\documentclass[twocolumn,epsfig,superscriptaddress,prb]{revtex4-2}

\usepackage{graphicx}
\usepackage{amssymb}
\usepackage{amsmath}
\usepackage{hyperref}
\usepackage[commandnameprefix=ifneeded]{changes}
\usepackage{mathtools,nccmath}
\usepackage{ragged2e} 
\usepackage{stackengine}
\usepackage{placeins}
\usepackage[percent]{overpic}
\usepackage{pgfplots}
\usepackage{tabularx}
\usepackage[labelformat=simple]{subcaption}
\usepackage{float}

\DeclareCaptionJustification{justified}{\justifying}
\definechangesauthor[name=Marcel, color=blue]{M}

\pgfplotsset{compat=1.8}
\DeclareMathAlphabet{\mathpzc}{OT1}{pzc}{m}{it}

\begin{document}
\title{Diode effect in microwave irradiated Josephson junctions with Yu-Shiba-Rusinov states}

\author{Aritra Lahiri}
\email[]{aritra.lahiri@mpl.mpg.de}
\affiliation{Institute for Theoretical Physics and Astrophysics,
University of W\"urzburg, D-97074 W\"urzburg, Germany}

\author{Marcel Pol\'ak}
\email[]{marcel.polak@uni-wuerzburg.de}
\affiliation{Institute for Theoretical Physics and Astrophysics,
University of W\"urzburg, D-97074 W\"urzburg, Germany}

\author{Bj\"orn Trauzettel}
\affiliation{Institute for Theoretical Physics and Astrophysics, University of W\"urzburg, D-97074 W\"urzburg, Germany}
\affiliation{W\"urzburg-Dresden Cluster of Excellence ct.qmat, Germany}
\date{\today}

\begin{abstract}
{We investigate the critical current in microwave-irradiated Josephson junctions hosting Yu-Shiba-Rusinov states due to magnetic impurities. 
Under two conditions, namely, (i) the breaking of particle-hole symmetry in the normal sense by non-zero potential scattering, and (ii) the breaking of inversion symmetry either by unequal magnitudes of potential scattering and/or magnetic moments, microwave irradiation induces an additional phase-independent contribution to the current. This leads to asymmetric critical currents for opposite current polarities, an effect absent in the same junction without microwave irradiation. The asymmetry is highly tunable via the microwave amplitude and frequency, and we may even achieve perfect asymmetry where the critical current vanishes for one polarity, akin to a perfect diode. While Yu-Shiba-Rusinov states provide the ideal platform for a pronounced asymmetry, we find that as long as the two conditions (i) and (ii) above are met, our proposal does not necessarily depend upon them.}
\end{abstract}
\maketitle

\section{Introduction}
The Josephson effect, describing the coherent flow of Cooper pairs across a junction between two superconductors, furnishes a fundamental signature of phase coherence in superconductors. In equilibrium, a minimal description yields a current-phase relation (CPR) of the form $I=I_c\sin(\phi)$, where $\phi$ is the Josephson phase. In conventional Josephson junctions (JJs), time-reversal and inversion symmetries typically enforce the antisymmetry $I(\phi)=-I(-\phi)$. Consequently, the critical current assume the same magnitude for both current polarities. However, in the absence of these symmetries, the CPR is not bound by the antisymmetry with respect to $\phi$, resulting in non-reciprocal critical currents characteristic of a diode effect (DE)~\cite{Hu2022,Chen2018,Davydova2022,Zhang2022,Nadeem2022,He2022,Misaki2021,Tanaka2022,Jiang2022,Wang2025,Wang2025a,Wang2025b}. Indeed, several experiments have reported a DE in JJs, typically requiring external magnetic fields to break time-reversal symmetry~\cite{Baumgartner2021,
Pal2022,Diezmerida2023,Wu2022,Bocquillon2017,Bauriedl2022,Jeon2022,Turini2022,Gupta2023,Chiles2023}. 

Moreover, there has been a concerted effort into realizing a ``field-free" Josephson DE, which exhibits non-reciprocal transport characteristics even in the absence of an external magnetic field~\cite{Wu2022,Kokkeler2022,Zhao2023,Zhang2024,Cheng2024,Banerjee2024,Ma2025,Nagata2025}. To this end, recent works have explored JJs hosting magnetic impurities/adatoms on a single lead as an avenue for time-reversal symmetry breaking~\cite{Trahms2023,Steiner2023,Ghosh2024,Trahms2025}. Such magnetic impurities generically couple to superconducting electrons by a combination of a spin-exchange, and a potential scattering term, leading to sub-gap Yu-Shiba-Rusinov (YSR) states~\cite{Yu1965,Shiba1968,Rusinov1969}. In Refs.~\cite{Trahms2023,Steiner2023,Ghosh2024,Trahms2025}, the authors investigated the role of broken particle-hole symmetry in the normal-state sense (PHN), amounting to an asymmetry of the normal-state density of states about the Fermi level. They showed that PHN symmetry breaking can generate a non-reciprocal dissipative current, leading to a DE in the retrapping current of current-biased Josephson junctions. Notably, this mechanism does not require breaking time-reversal symmetry. This contrasts with the more commonly studied DE in the switching current, which instead originates from non-reciprocity in the critical currents and therefore requires time-reversal symmetry breaking.

\begin{figure}
\begin{subfigure}[b]{0.95\columnwidth}
\caption{}
\includegraphics[width=0.75\linewidth]{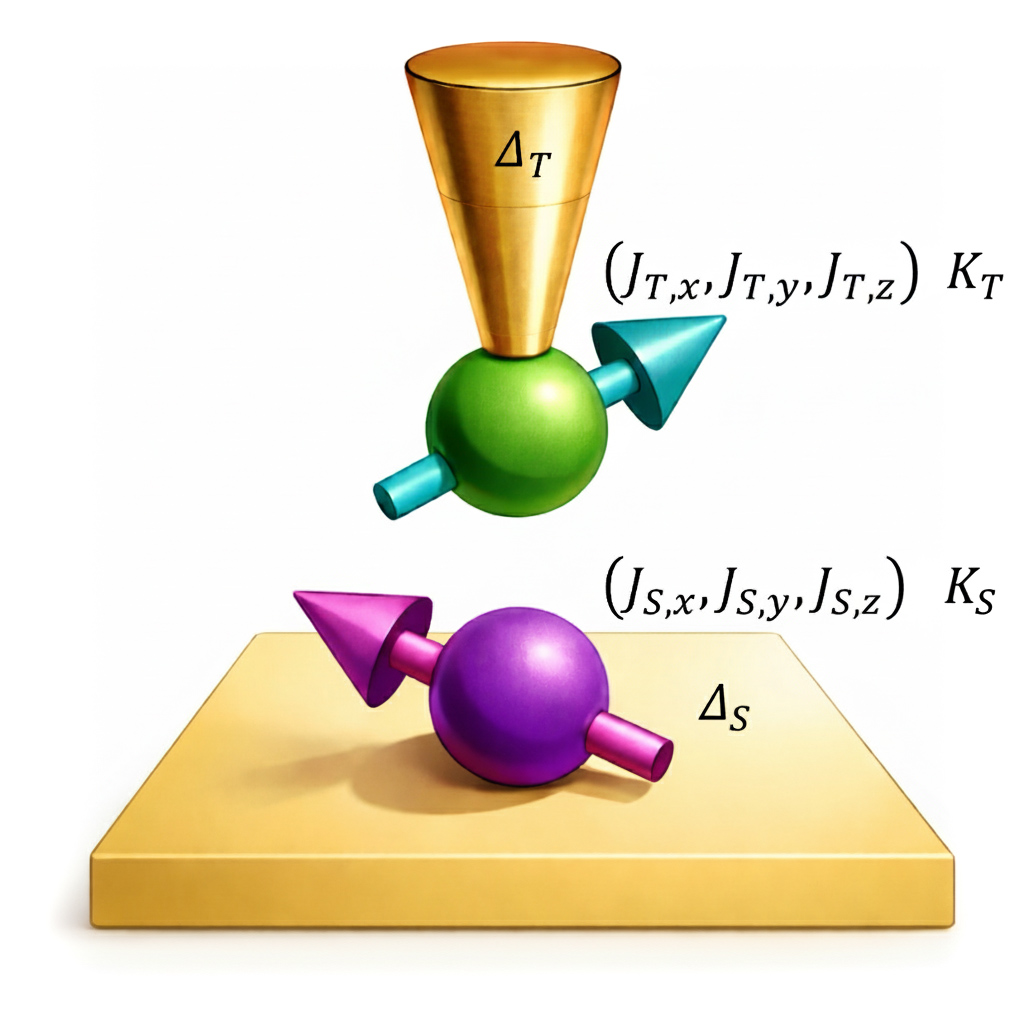}
\end{subfigure}
\begin{subfigure}[b]{0.99\columnwidth}
\caption{}
\includegraphics[width=0.8\linewidth]{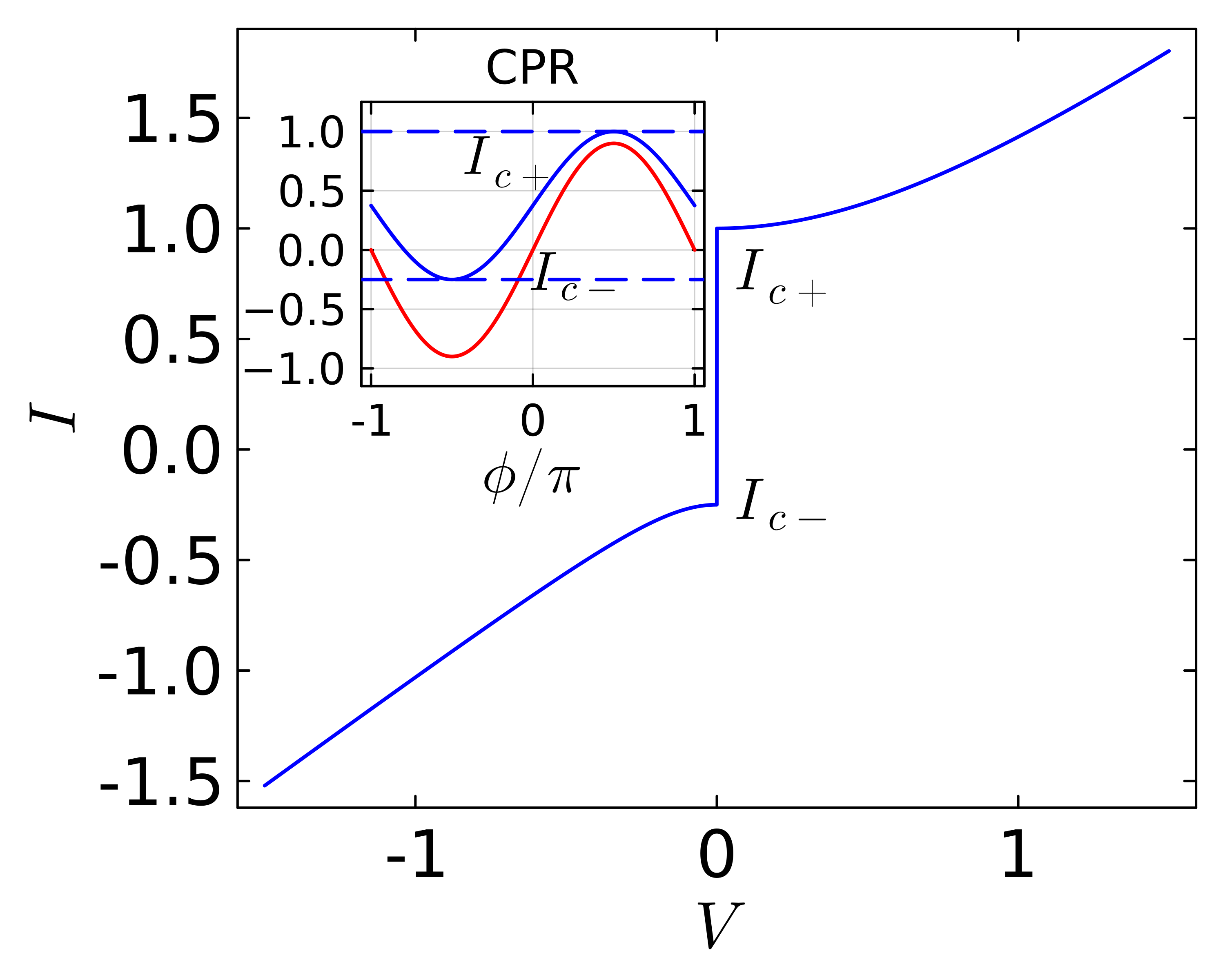}
\end{subfigure}
\caption{(a) Illustration of a JJ with arbitrarily oriented spins on the superconducting tip ($T$) and substrate ($S$), with gaps $\Delta_T$ and $\Delta_S$. $J_{T/S,k}$ and $K_{T/S}$ denote the magnetic moment and potential-scattering terms. For simplicity, we set $\Delta_T=\Delta_S$. (b) Illustration of the diode effect. The inset shows the CPR (arbitrary units) without (red) and with (blue) microwave radiation at zero voltage. While the former is centered at zero, the microwave-induced phase-independent contribution shifts it upward, resulting in two different effective critical currents $I_c^\pm$. Note that $I_c^+=I_c^-$ in the absence of microwaves. The main panel shows the current-voltage characteristics (arbitrary units). For currents in the range $[I_c^-, I_c^+]$, the JJ remains voltage-free, while outside this range it switches to a finite-voltage state. We only show the switching branch of the background current, leaving out the Shapiro steps and the retrapping current. }
\label{Fig1}
\end{figure}

In this work, we consider JJs hosting magnetic impurities, and hence YSR states, on both superconducting leads, driven by microwave radiation in the absence of an explicit dc voltage bias. We show  that when PHN is broken by non-zero potential scattering with the magnetic impurities in at least one of the two leads, and inversion symmetry is broken by an asymmetry in the magnitude of the magnetic moments and/or the potential scattering between the two leads, microwave driving induces a phase-independent contribution to the supercurrent. As shown in Fig.~\ref{Fig1}(b), this term shifts the CPR, effectively resulting in an asymmetry between the critical currents for opposite current polarities. Remarkably, this feature appears only under microwave driving, a purely phase- or current-biased JJ in the absence of microwaves lacks the phase-independent contribution and therefore exhibits equal critical currents in both directions. We further demonstrate that this phase-independent current, which is the origin of the DE, arises from the combined effect of inversion and PHN symmetry breaking. As a result, although the microwave radiation excites quasiparticles with equal probability in both directions, the broken symmetries render the transport non-reciprocal, producing a net microwave-induced current. The asymmetry in the critical currents is tunable by adjusting the radiation frequency and amplitude. In particular, we find that perfect asymmetry may be achieved, corresponding to an ideal diode, where the JJ sustains a finite critical current in one direction at zero voltage, while the critical current in the opposite direction vanishes.

The rest of this manuscript is organized as follows: In Sec. II, we elucidate the key concepts underpinning our proposed DE. Subsequently, we describe our microscopic Floquet-Keldysh formalism in Sec. III, followed by the numerical results in Sec. IV, exploring in detail the microwave-assisted DE and the dependence of YSR couplings. Finally, we conclude in Sec. V.

\section{Phenomenology}
\label{phenom}
In this section, we provide an intuitive picture of the microwave-incuded DE. First, we start with a simple description of the effects of microwave radiation on the Josephson current. In the presence of microwave radiation, which imposes an ac voltage $V_{\text{ac}}\cos(\omega_r t)$~\cite{Baronebook,Cuevas2002,Chauvinthesis,Shapiro1963,Kot2020,Siebrecht2023}, the Josephson phase is obtained as (we use $\hbar=1$ throughout)
\begin{align}
\phi(t) = \phi + \underbrace{(2eV_{\text{ac}}/\omega_r)}_{\coloneqq \alpha}\sin(\omega_r t),\label{Jphase}
\end{align}

We start with the ``adiabatic" approximation (AA) \cite{Baronebook,Bergeret2010}. The starting point in this approximation is the 
expression for the equilibrium Josephson current in the absence of microwaves
\begin{align}
I=\sum_{n\geq 1} \hat{I}_{\sin}^{(n)}\sin(n\phi).
\end{align} 
A microscopic analysis shows that the harmonics $\hat{I}^{(n)}$ progressively decrease with increasing $n$, $|\hat{I}^{(n)}|>|\hat{I}^{(n+1)}|$, 
and contribute noticeably only at high transparencies. Within the AA, we substitute the Josephson phase of Eq.~\eqref{Jphase} in the equilibrium 
CPR, resulting in 
\begin{align}
I=& \sum_{n\geq 1} \sum_{a=-\infty}^\infty \underbrace{ \hat{I}_{\sin}^{(n)}J_a(n\alpha) }_{\textstyle  I_{\sin}^{(n),a}(\alpha) } \sin(n\phi + a\omega_r t). \label{IAA1}
\end{align}
where $J_n$ denotes the $n^\mathrm{th}$ order Bessel function of the first kind and we define the dressed amplitudes $I_{\sin}^{(n),a}(\alpha)$ incorporating the effect of the microwaves. This approximation works only for slowly varying $\phi(t)$, which permits us to neglect the intrinsic frequency dependence of $\hat{I}_{\sin}^{(n)}$ and replace them with their zero-frequency values \cite{Baronebook,Bergeret2010}. This is a crucial point, and we shall revisit this matter soon below. The DC supercurrent follows as
\begin{align}
I=& \sum_{n\geq 1}  I_{\sin}^{(n),0}(\alpha) \sin(n\phi), \label{IDCAA}
\end{align}
For the transparencies considered in this work, we do not find any significant higher harmonic contributions in our microscopic numerical calculations. Thus, the critical current is obtained as $I_c=I_{\sin}^{(1),0}(\alpha)=\hat{I}_{\sin}^{(1)}J_0(\alpha)$. Uptil this point, there is no DE as the response is symmetric for positive and negative current polarities.

However, this conclusion changes on going beyong the AA. As mentioned above, within the AA, we had neglected the intrinsic frequency dependence of $\hat{I}_{\sin}^{(1)}$. This is typically reasonable in clean Josephson junctions with no sub-gap bound states, which results in an approximately frequency-independent $\hat{I}_{\sin}^{(1)}$ for frequencies smaller than $2\Delta$~\cite{Baronebook,Werthamer1966,Larkin1967}. In the presence of low-energy YSR states, $\hat{I}_{\sin}^{(1)}$ exhibits sharp resonances which break this approximation. However this, on its own, is not sufficient to obtain a DE. Remarkably, to this end, we discover that microwave-driven microscopic electronic excitations generate two additional contributions to the DC current. Under two conditions, we obtain a term proportional to $\cos(\phi)$ as well as a phase-independent constant contribution: (i) when the YSR potential scattering terms are non-zero, and (ii) concomitantly inversion symmetry is broken by an asymmetry in the magnitudes of the potential scattering and/or the magnetic moments in the two leads. This results in a DC current
\begin{align}
I= &I_{\sin}^{(1),0}(\alpha)\sin(\phi) + I_{\cos}^{(1),0}(\alpha)\cos(\phi) + I_{\text{con}}^{(1),0}(\alpha), \label{IDCmicro}
\end{align}
where $I_{\sin}^{(1),0}(\alpha)=\hat{I}_{\sin}^{(1)}J_0(\alpha)$ as before. However, $I_{\cos}^{(1),0}(\alpha)$ and $I_{\text{con}}^{(1),0}(\alpha)$ bear no such simple relation as, in the absence of radiation, $\hat{I}_{\cos}^{(1),0}=0$ and $\hat{I}_{\text{con}}^{(1),0}=0$. Hence, the frequency dependence of $I_{\cos}^{(1),0}(\alpha)$ and $I_{\text{con}}^{(1),0}(\alpha)$ is a purely microscopic phenomenon, beyond the scope of the AA. While the term varying as $\cos(\phi)$ is still incapable of realizing a DE, the phase-independent term $I_{\text{con}}^{(1),0}(\alpha)$ does the trick. It shifts the CPR for all $\phi$, and thus, results in a different critical current for positive and negative current polarities.

Physically speaking, the phase-independent (as well as the $\cos(\phi)$) term arises when the microwaves produce real (as opposed to virtual) excitations between the YSR states, resulting in a dissipative current. Indeed, in the absence of YSR states, such contributions arise only for microwave frequencies exceeding $2\Delta$, the excitation gap of a clean superconductor. The presence of low-energy YSR states substantially lowers this threshold and simultaneously provides a mechanism to break inversion and PHN symmetries, which are necessary for realizing a DE. In order to see this, we look at the microscopic expression for the phase-independent DC tunnel current in the presence of microwaves. Starting with the expression for the tunnel DC current-voltage characteristics (IVC)~\cite{Huang2021}, $I_{\mathrm{dc},V}$, the Tien-Gordon theory provides the tunnel current for microwave-irradiated junctions~\cite{Tien1963,Falci1991,Baronebook,Safi2010,Safi2019}
\begin{align}
&I_{\text{con}}
=
\sum_{n=-\infty}^{\infty}
J_n^2\Big(\frac{\alpha}{2}\Big)
\, I_{\mathrm{dc},V}(n\omega_r)\nonumber\\
&=\sum_{n>0}
J_n^2\Big(\frac{\alpha}{2}\Big)
\, \big[ I_{\mathrm{dc},V}(n\omega_r) + I_{\mathrm{dc},V}(-n\omega_r) \big],\label{ITG}
\end{align}
which is the phase-independent contribution. For typical junctions with both inversion and PHN symmetries, the current response to DC voltages of opposite polarities is reciprocal, $I_{\mathrm{dc}, V}(-V)=-I_{\mathrm{dc}, V}(V)$, implying that $I_{\text{con}}$ vanishes. This is reasonable as microwaves excite tunneling quasiparticles equally in both directions, which are equal and opposite in sign, causing the current to vanish. However, for junctions breaking inversion and PHN symmetries, $I_{\mathrm{dc}, V}(-V)\neq -I_{\mathrm{dc}, V}(V)$~\cite{Steiner2023}. Consequently, $I_{\text{con}}\neq 0$, as even though the microwave excitation is symmetric, the non-reciprocity of the current prevents an exact cancellation.

Let us now elaborate a bit on the requirements for breaking inversion and PHN symmetries: (a) First, PHN symmetry is broken when at least one of the two leads has a non-zero potential scattering associated with the magnetic impurity~\cite{Steiner2023,Salkota1997,Flette1997,Balatsky2006,Villas2020}. On flipping the sign of the voltage, the transport pathways are mirrored in frequency about the Fermi level. Breaking PHN with the potential scattering effectively shifts the chemical potential away from the center of the band, resulting in an asymmetric excitation spectrum and density of states about the Fermi level. This makes the current transport amplitudes different for opposite voltage polarities, as we show in Appendix~\ref{appA} for a tunnel JJ. (b) Second, inversion symmetry is broken when the \emph{magnitudes} of the potential scattering and/or the magnetic moment of the magnetic impurities in the two leads differ. While the former is obvious, the latter has a few subtleties. When both magnetic moments have the same magnitudes, simply rotating one magnetic impurity relative to the other, in the absence of spin-orbit coupling, does not break inversion symmetry. We explain it in detail below, after introducing the technical details. This follows from the fact that the spin-space is entirely independent of the real-space (in the absence of spin-orbit coupling), and hence the JJ is inversion symmetric irrespective of the orientation of the magnetic impurities as long as the magnetic moments of the two leads have the same magnitude. In this case, we do not expect a microwave-induced DE.

\section{Microscopic Model} 
\label{micromodel}
We employ the Floquet-Keldysh formalism following Refs.~\cite{Cuevas1996,Cuevas2002,Lahiri2025,Lahiri2026}. Nevertheless, 
we describe below the essential aspects for a self-contained discussion. As illustrated in Fig.~\ref{Fig1}, we consider a Josephson point contact, modelling our junction with single-channel s-wave BCS Hamiltonians with a spin at the apex of each lead near the junction~\cite{Chakraborty2020,Villas2020}. Defining the extended Nambu spinor $\check{c}=[c_{\uparrow}\ c_{\downarrow}\ c^\dagger_{\uparrow}\ c^\dagger_{\downarrow}]^T$~\cite{Beenakker2015}, 
 and the Pauli matrices in the particle-hole ($\tau$) and spin ($\sigma$) space, the Hamiltonian is given by $H=\sum_{\gamma=T/S}H_\gamma+H_{\mathcal{T}}$, where the lead [$\gamma=T(S):$ tip(substrate)] Hamiltonian is~\cite{Huang2021}

\begin{align}
H_{\gamma}& =  \nonumber\\
\sum_{j\in\gamma} \ & \check c_{\gamma j}^\dagger
\Big[ -\mu\ \tau_z\!\otimes\!\sigma_0
+ \Delta\ \big( \tau_+\!\otimes\! i\sigma_y
+ \tau_-\!\otimes\! (-i)\sigma_y \big) \Big] \check c_{\gamma j} \nonumber \\
+\ & \check c_{\gamma j}^\dagger\big[-\zeta\  \tau_z\!\otimes\!\sigma_0\, \big] \check c_{j+1} + \check c_{j+1}^\dagger\big[-\zeta\  \tau_z\!\otimes\!\sigma_0\, \big] \check c_{\gamma j}\nonumber\\
+\ &\check c_{\gamma 1}^\dagger
\bigg[ \sum_{k=x,y,z}\!\!\!\! J_{\gamma,k}\ \big(\tau_u\!\otimes\!\sigma_k - \tau_d\!\otimes\!\sigma_k^*\big)
+ K_{\gamma}\ \tau_z\!\otimes\!\sigma_0 \bigg] \check c_{\gamma 1}
\label{Ham}
\end{align}
with $\tau_{u/d}=(\tau_0\pm\tau_z)/2$ and $\tau_{\pm}=(\tau_x\pm i\tau_y)/2$, and the tunnel Hamiltonian is
\begin{subequations}
\begin{align}
H_{\mathcal{T}}=&\ \check{c}_{T1}^\dagger\big[W(t)\ \tau_u \otimes  \sigma_0 - W^*(t)\ \tau_d \otimes  \sigma_0 \big] \check{c}_{S1} + h.c.,\\
W(t)=&\ -\mathcal{T}e^{i\frac{\phi(t)}{2}}.
\end{align}
\end{subequations}
Here, $\mathbf{J}_{\gamma}=[J_{\gamma,x},\ J_{\gamma,y},\ J_{\gamma,z}]$ denotes the classical magnetic moment of the magnetic impurity in the lead $\gamma$, while $K_\gamma$ denotes the corresponding potential scattering term. We have assumed spin-conserving tunneling. Note that we employ a gauge wherein the phases of the superconducting order parameters have been transferred to the tunnel coupling~\cite{Cuevas1996}. 

In microwave irradiated JJs, using Eq.~\eqref{Jphase}, the tunnel amplitude $W(t)$ equals
\begin{align}
W(t)=&\sum_{n=-\infty}^\infty \underbrace{-\mathcal{T}e^{-i\phi}J_n\Big(\mfrac{\alpha}{2}\Big)}_{W_{n}} \label{Wmicro}
e^{-in\omega_r t},
\end{align}
revealing that energy is transferred in units of $\omega_r$. For a numerical solution, we formulate an effective theory for the apex of the two superconducting leads, integrating out the rest of the semi-infinite superconducting reservoirs~\cite{Lahiri2025}. In this setup, for brevity, we simply denote the apex sites of the $T(S)$ leads by $j=T(S)$. The tunneling self-energy is given by
\begin{subequations}
\begin{align}
\Sigma^{r/a}_{\mathcal{T},ST/TS}(t)&=H_{\mathcal{T},ST/TS},\\
\Sigma^{r/a}_{\mathcal{T},ST;mn}&=W_{(m-n)}\tau_u\! \otimes \! \sigma_0 - W_{-(m-n)}^*\tau_d\! \otimes \! \sigma_0,\\
\Sigma^{r/a}_{\mathcal{T},TS}&=\Sigma^{r/a\dag}_{\mathcal{T},ST}
\end{align} 
\end{subequations}
Note that $\Sigma_{\mathcal{T}}^<=0$~\cite{Cuevas1996}.
Therefore, the two-time retarded and advanced Green's functions may be written as~\cite{Lahiri2026}

\begin{align}
G^{r/a}(t,t')=&\sum_{q}\! {\int\limits_{-\infty}^{\infty}} \! \frac{d\omega}{2\pi} \overbrace{G^{r/a}
\big(\omega+q\omega_r,\omega\big)}^{G^{r/a}_{q}(\omega)} e^{-i(\omega+q\omega_r ) t+i \omega t'}.\label{Grep}
\end{align}
In this representation, the Dyson equations for the retarded and advanced components become
\begin{align}
G^{r/a}_{q}(\omega)=&g^{r/a}_{0}(\omega)\delta_{q,0}+\sum_{n}g^{r/a}_{q}(\omega)\Sigma^{r/a}_{qn}
G^{r/a}_{n}(\omega),\label{RADyson}
\end{align}
where all quantities are matrices in Nambu space, and the bare Green's function $g_{q}(\omega)=g(\omega+q\omega_r)$ 
is defined in the absence of tunneling. Apart from the tunnling self-energy, we also have the reservoir self-energy, $\Sigma^{r/a/<}_{\text{res\deleted{.}};pq,mn}=\zeta^2\tau_3 g_b^{r/a/<} \tau_3 \delta_{p,m}\delta_{q,n}$, where $g_b$ is the boundary Green's function of the semi-infinite superconducting chains~\cite{Samanta1998}. It represents the macroscopic superconducting reservoirs connected to the apexes of the superconducting leads containing the magnetic impurities. For numerical convenience, we employ single site leads, containing only the apexes. Lastly, we include the broadening self-energy $\Sigma^{r/a}_{\Gamma;pq,mn}=\mp i(\Gamma/2) \delta_{p,m}\delta_{q,n}$ and $\Sigma^{<}_{\Gamma;pq,mn}=i\Gamma f(\omega) \delta_{p,m}\delta_{q,n}$, where $f(\omega)$ is the 
Fermi function. It aids numerical convergence, and accounts for the lifetime arising from, e.g., relaxation to the quasiparticle 
continuum, electron-phonon interaction, etc~\cite{Lahiri2025}. Finally, the lesser Green's function, which is central to our 
calculation, is obtained as~\cite{Jauho1994,Keldysh1964,Stefanuccibook2013,Gonzales2020,Wimmerthesis,Gldecaynote}
\begin{align}
G^<_{m}(\omega)=& \sum_{n} G^{r}_{n+m}(\omega)\Sigma^<(\omega) {G^{r}}^\dagger_{n}(\omega),\label{LDyson}
\end{align}
where we have used $G^a(w,\omega+n\omega_r)={G^r}^\dagger(\omega+n\omega_r,w) \equiv 
{G^r_{n}}^\dagger$. 

Then, the current is obtained as~\cite{Lahiri2025,Cuevas2002},
\begin{subequations}
\begin{align}
I(t)=&\sum_{m}I_{m}e^{-im\omega_r t}\\
I_{m}=&\sum_{p}e \int_{-\infty}^{\infty} \frac{d\omega}{2\pi} \mathbf{tr}\big[ (\tau_3\otimes\sigma_0)\Sigma_{\mathcal{T},TS,m} 
G_{ST,p}^{<}(\omega)\nonumber\\
&-(T\leftrightarrow S)\big].\label{If}
\end{align}
\end{subequations}
The CPR is derived as the DC component, $I_0$ from Eq.~\eqref{If}, by varying the constant part of the Josephson phase, $\phi$, in Eq.~\eqref{Wmicro}.

\section{Results}

In this section, we explain the conditions to observe the DE within our microscopic model. Subsequently, we present our numerical results for the critical currents and the CPR, following the theory outlined in the previous sections.

We recall that the realization of the DE, characterized by unequal critical currents for positive ($I_c^+$) and negative ($I_c^-$) current polarities, requires two conditions to be satisfied: (a) PHN must be broken by a non-zero $K_\gamma$ in at least one of the two leads. We obtain explicit expressions for the tunnel current in App.~\ref{appA}, showing how the presence of PHN as well as inversion symmetries preclude a non-reciprocal response in the IVC, and hence, no microwave-induced DE. (b) Inversion symmetry must be broken, which can be achieved by $K_T\neq K_S$, and/or $|\mathbf{J}_T|\neq |\mathbf{J}_S|$. On that note, we can now clarify why merely rotating the magnetic impurities is insufficient to break inversion symmetry if they have the same magnitude. To see this, suppose $\mathbf{J}_T$ and $\mathbf{J}_S$ have the same magnitude $J$, but are not parallelly oriented due to a relative angle $\theta=\text{acos}(\mathbf{J}_T\cdot \mathbf{J}_S/(|\mathbf{J}_T||\mathbf{J}_S|))$. In the absence of spin-orbit coupling, we are free to rotate the spin-degrees of freedom of the fermions to our convenience. We choose $\mathbf{J}_T$ to lie along the $z-$axis, and let the $x-$axis be the transport direction directed from the tip to the substrate. $\mathbf{J}_S$ now makes an angle $\theta$ with the $z-$axis, and $\theta'=\text{acos}(\mathbf{J}_{S,x}/\sqrt{\mathbf{J}_{S,x}^2+\mathbf{J}_{S,y}^2})$ with the  \begin{figure*}
\begin{subfigure}[b]{0.31\linewidth}
\caption{$\ K_T/\Delta_T=K_S/\Delta_T=2$, $|\mathbf{J}_T|/\Delta_T=|\mathbf{J}_S|/\Delta_T=5$, $\theta=\pi/2$}\label{subfig:2e}
\includegraphics[width=0.999\columnwidth]{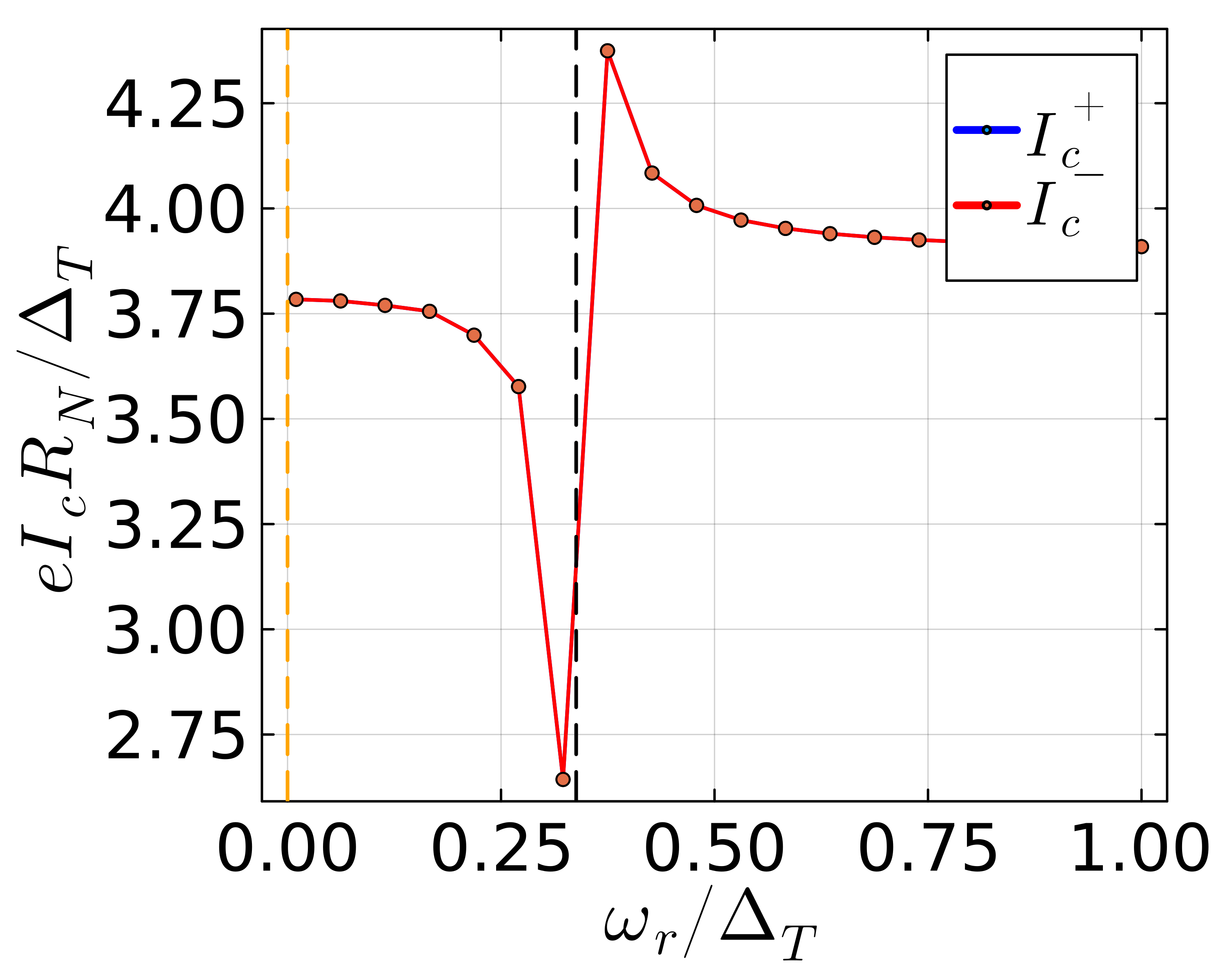}
\end{subfigure}
\begin{subfigure}[b]{0.31\linewidth}
\caption{}\label{subfig:2f}
\includegraphics[width=0.999\columnwidth]{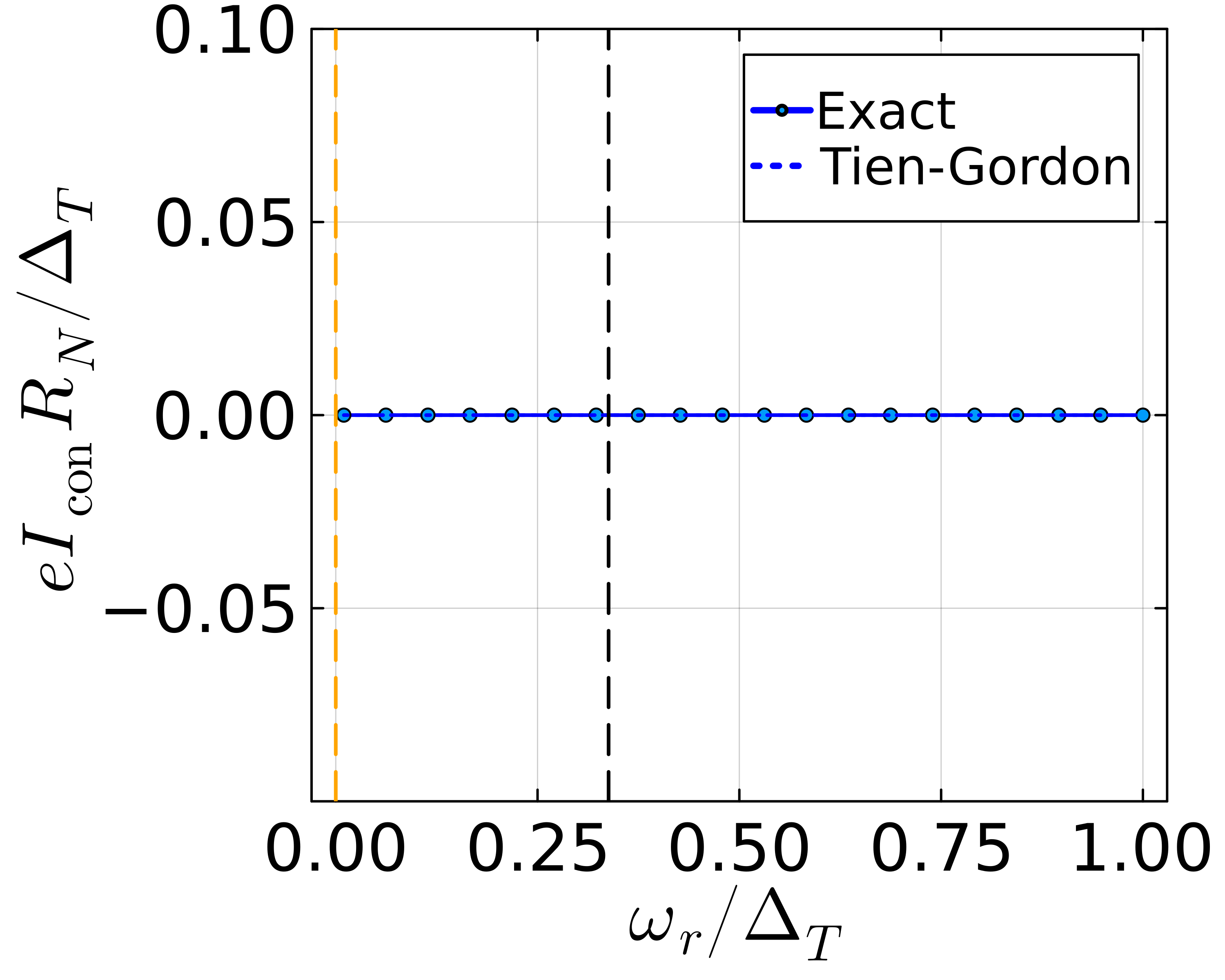}
\end{subfigure}
\begin{subfigure}[b]{0.31\linewidth}
\caption{}\label{subfig:2g}
\includegraphics[width=0.999\columnwidth]{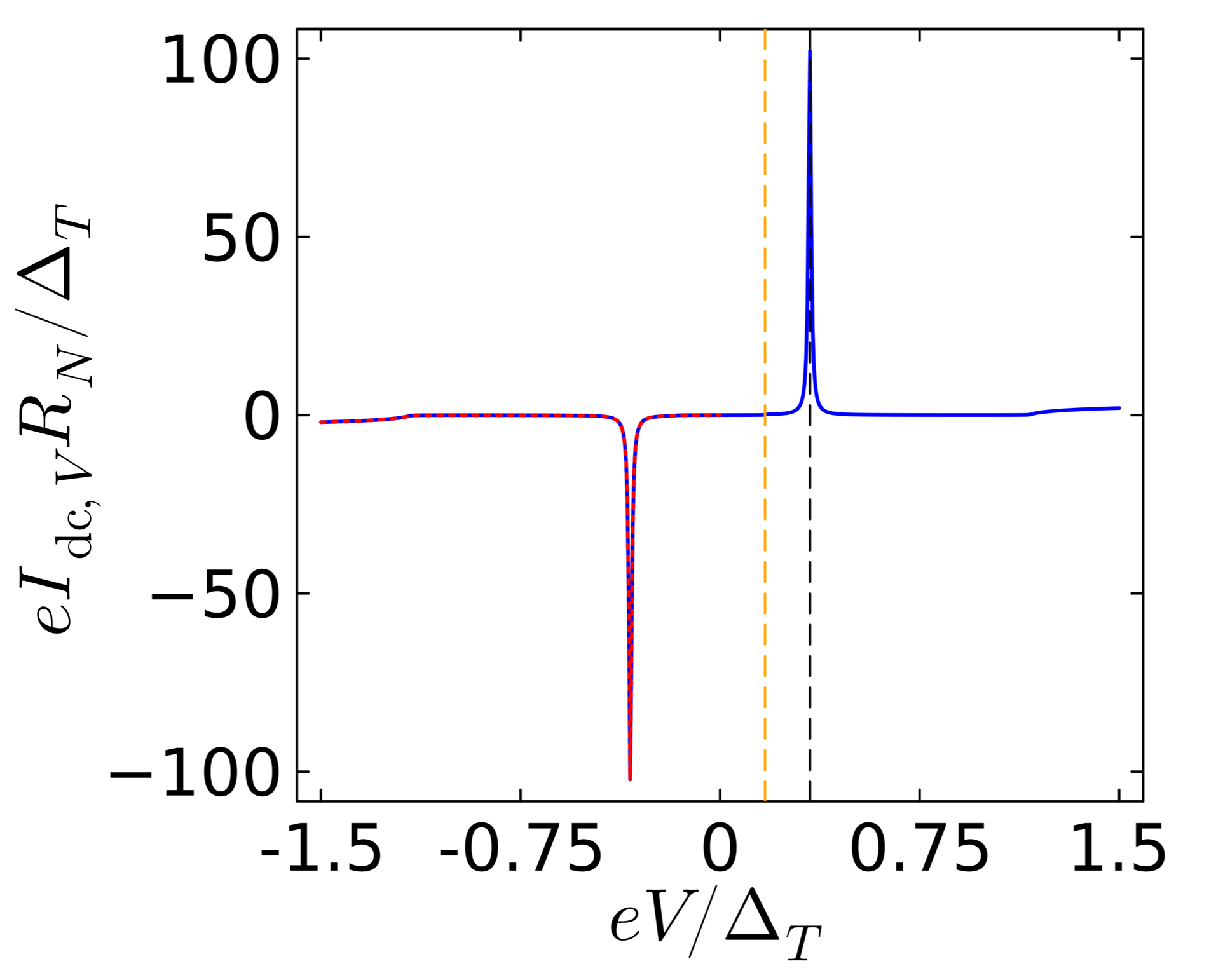}
\end{subfigure}
\begin{subfigure}[b]{0.31\linewidth}
\caption{$\ K_T/\Delta_T=3$, $K_S/\Delta_T=1$, $|\mathbf{J}_T|/\Delta_T=|\mathbf{J}_S|/\Delta_T=5$, $\theta=\pi/2$}\label{subfig:2h}
\includegraphics[width=0.999\columnwidth]{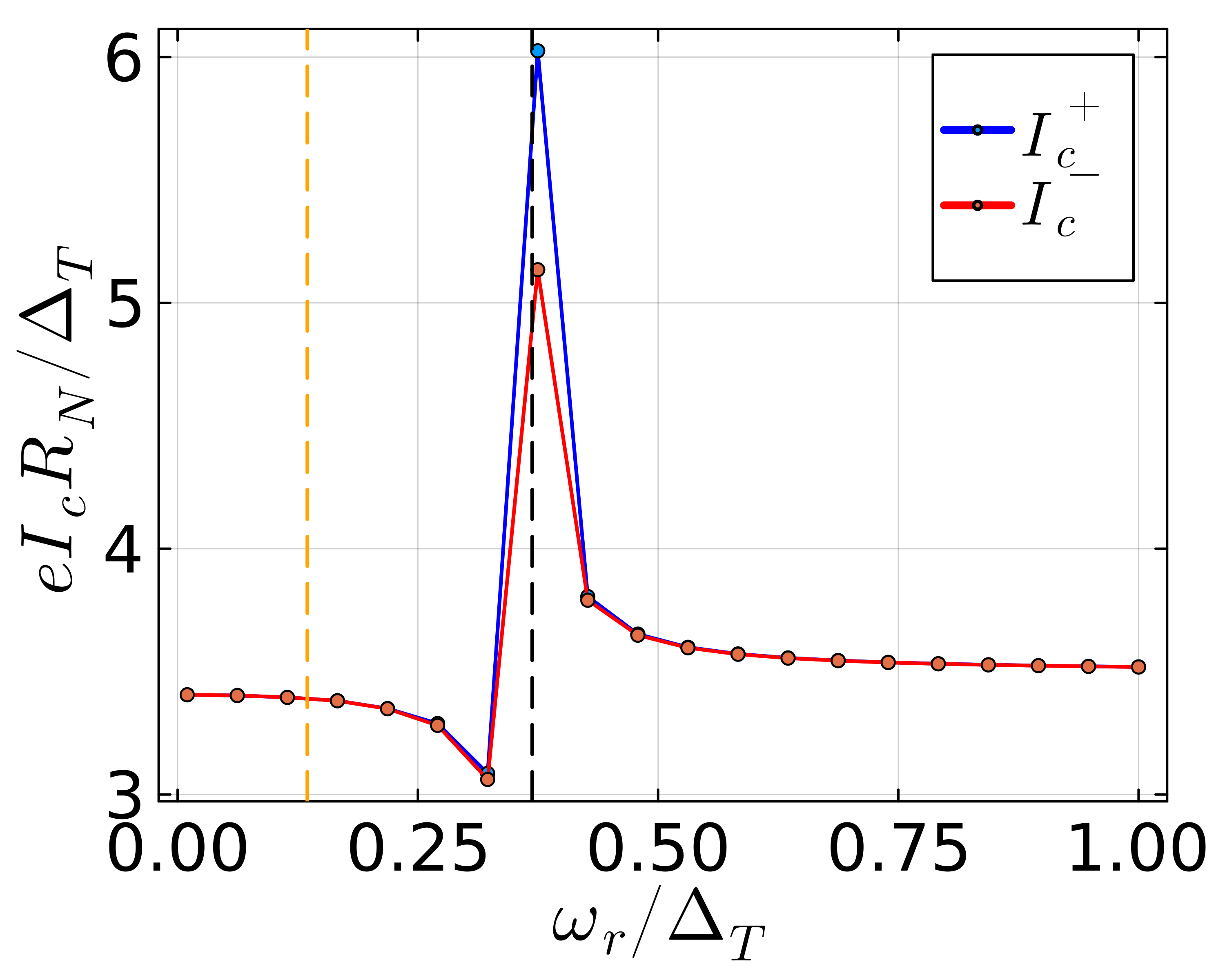}
\end{subfigure}
\begin{subfigure}[b]{0.31\linewidth}
\caption{}\label{subfig:2i}
\includegraphics[width=0.999\columnwidth]{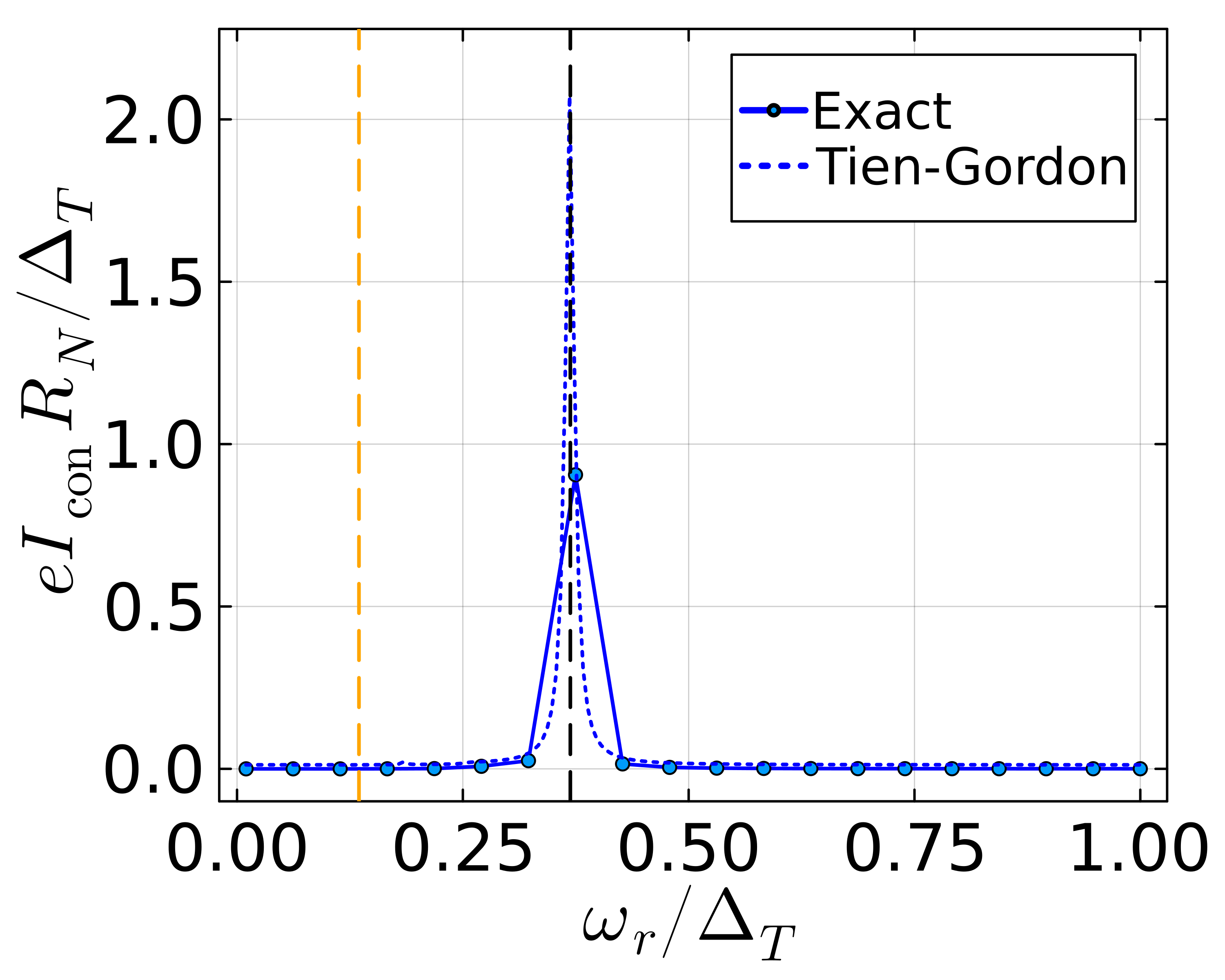}
\end{subfigure}
\begin{subfigure}[b]{0.31\linewidth}
\caption{}\label{subfig:2j}
\includegraphics[width=0.999\columnwidth]{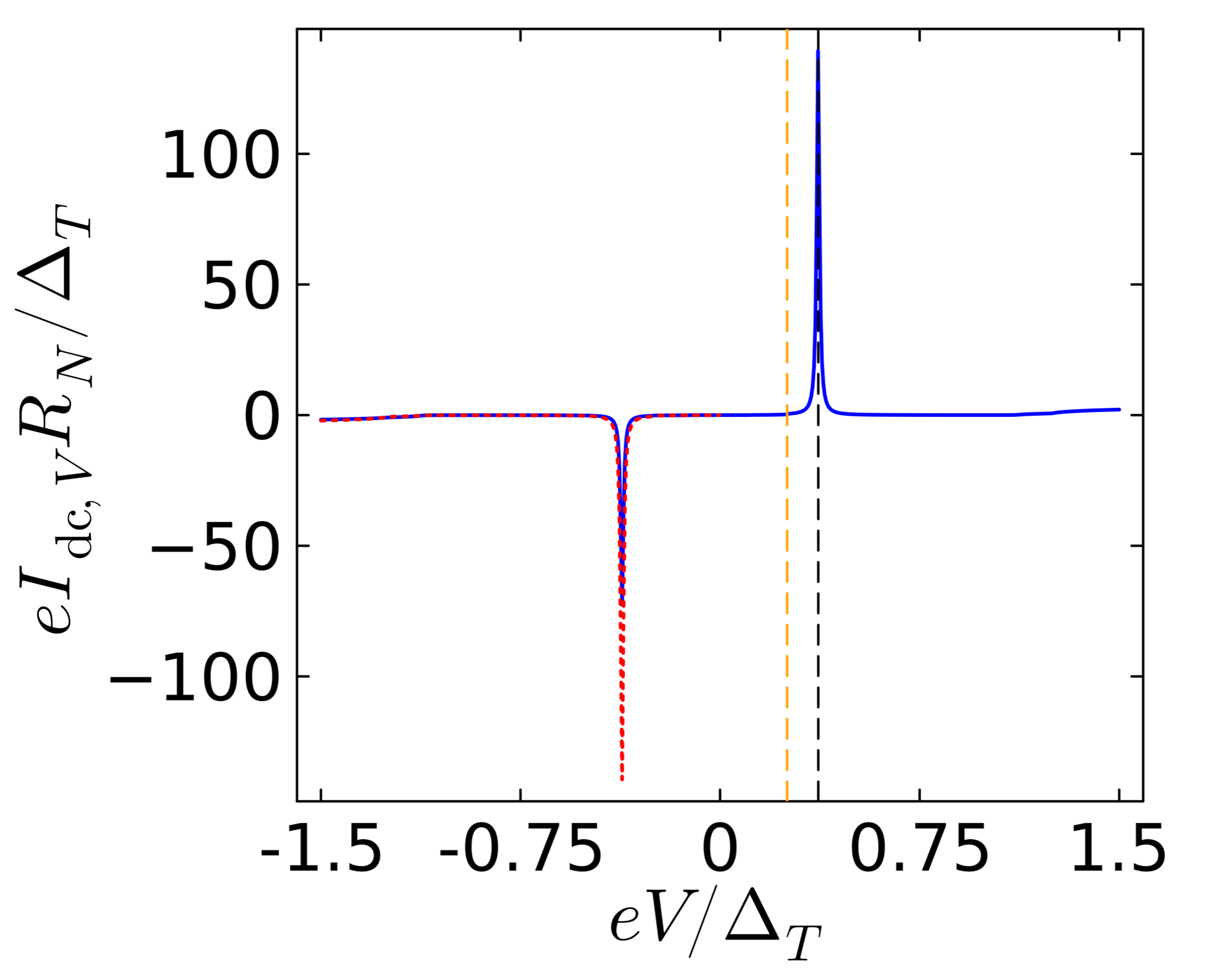}
\end{subfigure}
\begin{subfigure}[b]{0.31\linewidth}
\caption{$\ K_T/\Delta_T=2$, $K_S/\Delta_T=2$, $|\mathbf{J}_T|/\Delta_T=5$, $|\mathbf{J}_S|/\Delta_T=6$, $\theta=0$}\label{subfig:2h}
\includegraphics[width=0.999\columnwidth]{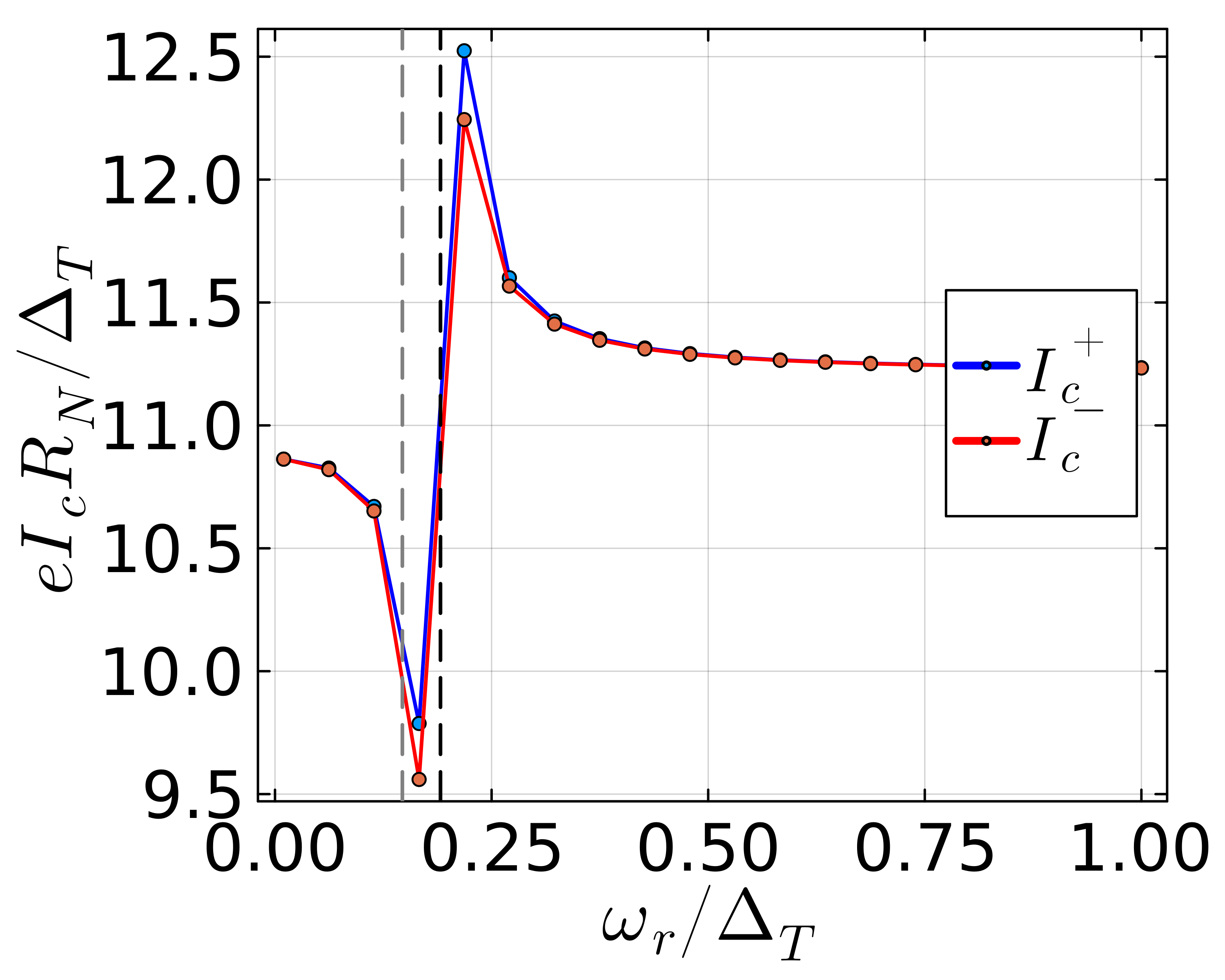}
\end{subfigure}
\begin{subfigure}[b]{0.31\linewidth}
\caption{}\label{subfig:2i}
\includegraphics[width=0.999\columnwidth]{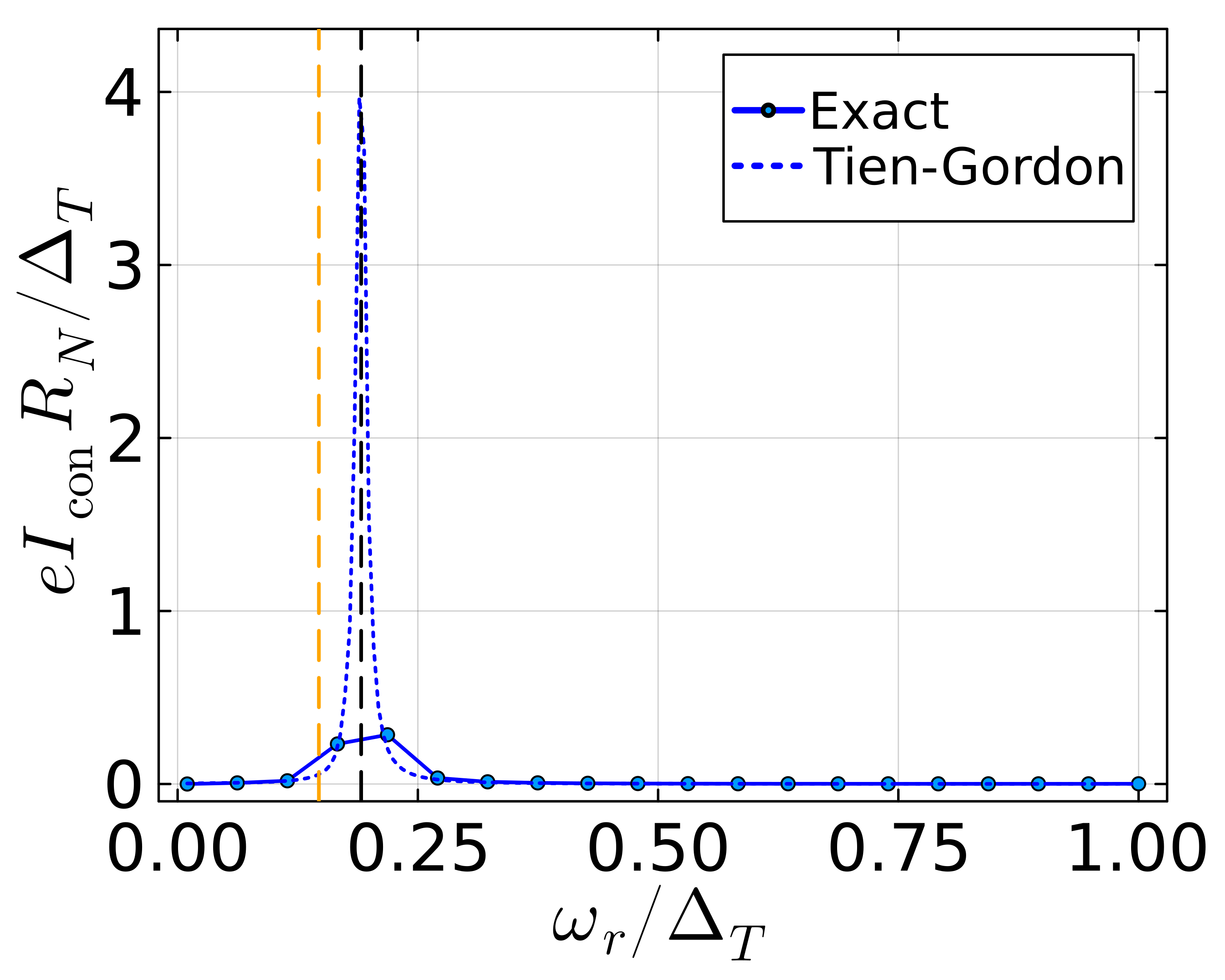}
\end{subfigure}
\begin{subfigure}[b]{0.31\linewidth}
\caption{}\label{subfig:2j}
\includegraphics[width=0.999\columnwidth]{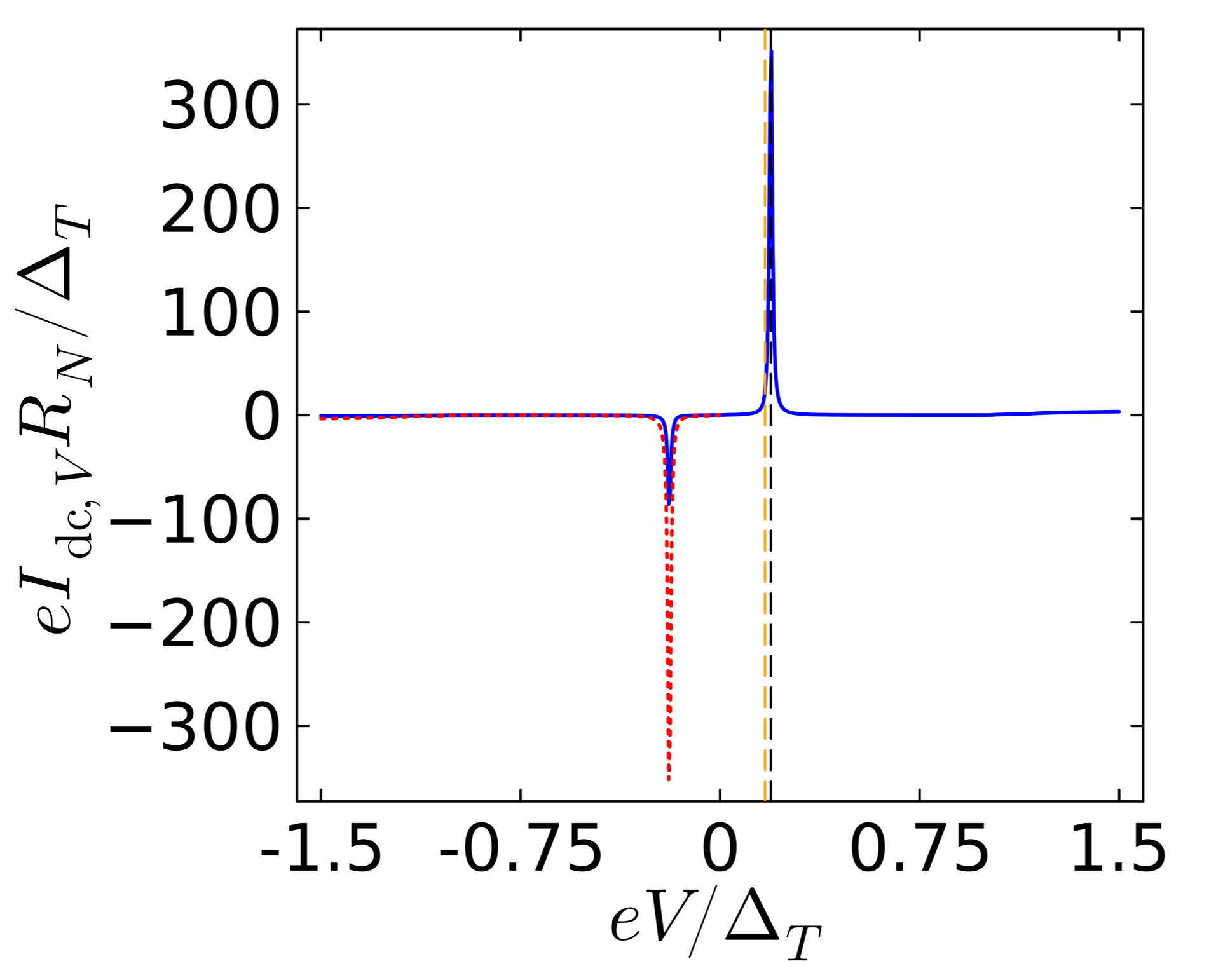}
\end{subfigure}
\caption{Numerically obtained currents for $\zeta=5\Delta_T=5\Delta_S$, $\mu=0$, $\mathcal{T}=0.002\zeta$ (normal-state transparency $\sim10^{-5}$~\cite{Cuevas1996}), $\Gamma=2\times10^{-3}\Delta_T$, and radiation strength $\alpha=0.5$. Currents are normalized by the normal-state resistance $R_N$. Left panels show the magnitudes of the critical currents $I_c^{\pm}$ as functions of $\omega_r$. Central panels display the corresponding $I_{\cos}$. Circular markers (joined by straight lines as a guide to the eye) represent the exact result, while the dotted lines indicate the corresponding Tien-Gordon prediction obtained from Eq.~\eqref{ITG}. The markers lie essentially on top of the dotted curve, demonstrating good agreement. Right panels show the IVC $I_{\mathrm{dc},V}$, where the dotted curve for negative voltages represents $-I_{\mathrm{dc},V}(-V)$; non-overlapping curves indicate a non-reciprocal response. Dashed lines mark the YSR transition energies $\epsilon_{0,T}+\epsilon_{0,S}$ (gray) and $|\epsilon_{0,T}-\epsilon_{0,S}|$ (green)~\cite{Huang2021,tdnote}, where $\epsilon_{0,T/S}$ denotes the YSR energy in lead $T/S$. Panels (a-c) correspond to an inversion-symmetric junction with $K_T=K_S$ and $|\mathbf{J}_T|=|\mathbf{J}_S|$ at a relative angle $\pi/2$, yielding $I_c^+=I_c^-$ and no DE. In (d-f), inversion symmetry is broken by $K_T\neq K_S$ with $\mathbf{J}_T=\mathbf{J}_S$, resulting in a DE driven by a finite phase-independent current. Panels (g-i) show the case $K_T=K_S\neq0$ with unequal magnetic moment magnitudes, which also breaks inversion symmetry and produces a DE.}
\label{Fig2} 
\end{figure*}
\begin{figure*}
\begin{subfigure}[b]{0.325\linewidth}
\caption{}\label{subfig:1cprb}
\includegraphics[width=0.999\columnwidth]{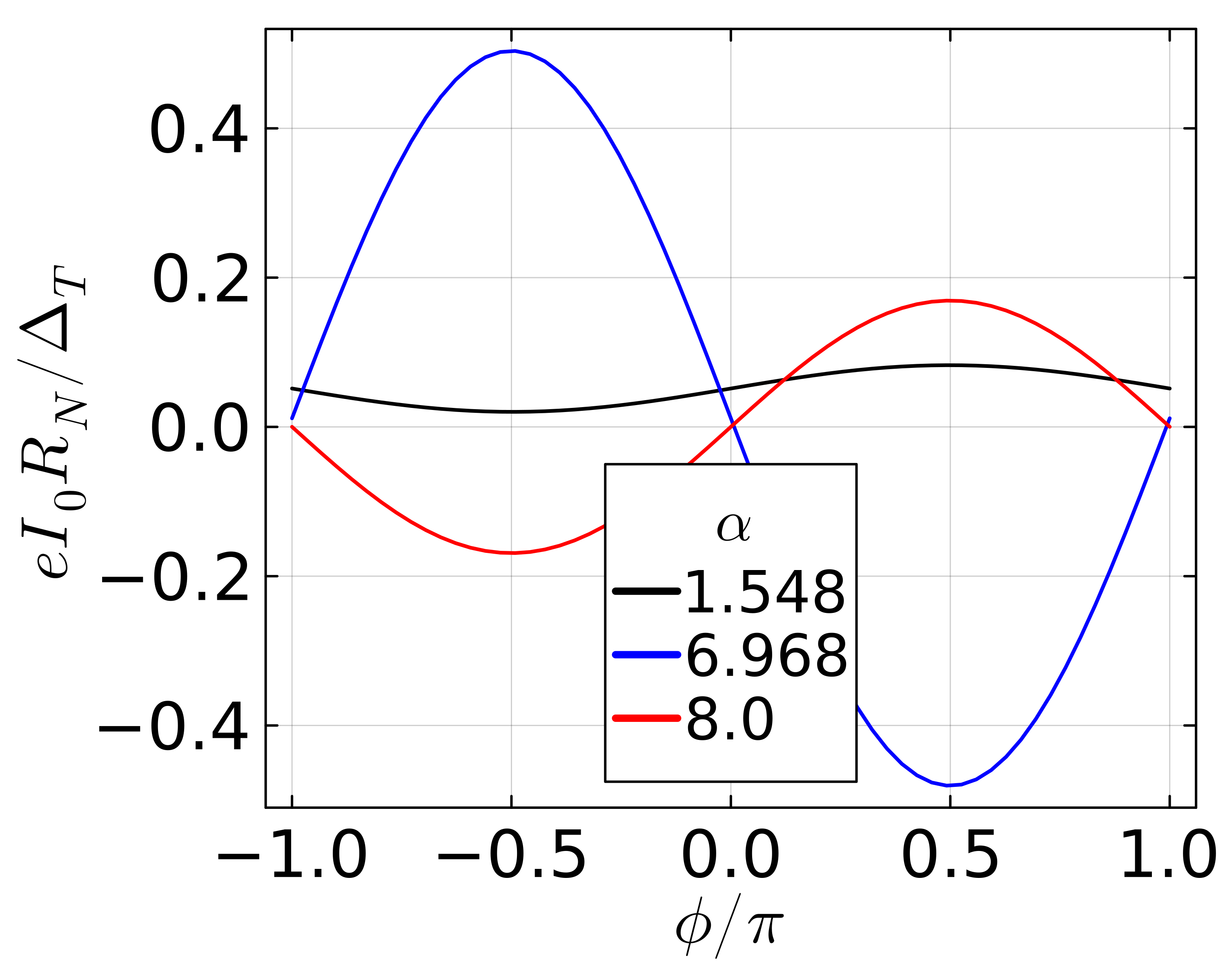}
\end{subfigure}
\begin{subfigure}[b]{0.325\linewidth}
\caption{}\label{subfig:1cprb}
\includegraphics[width=0.999\columnwidth]{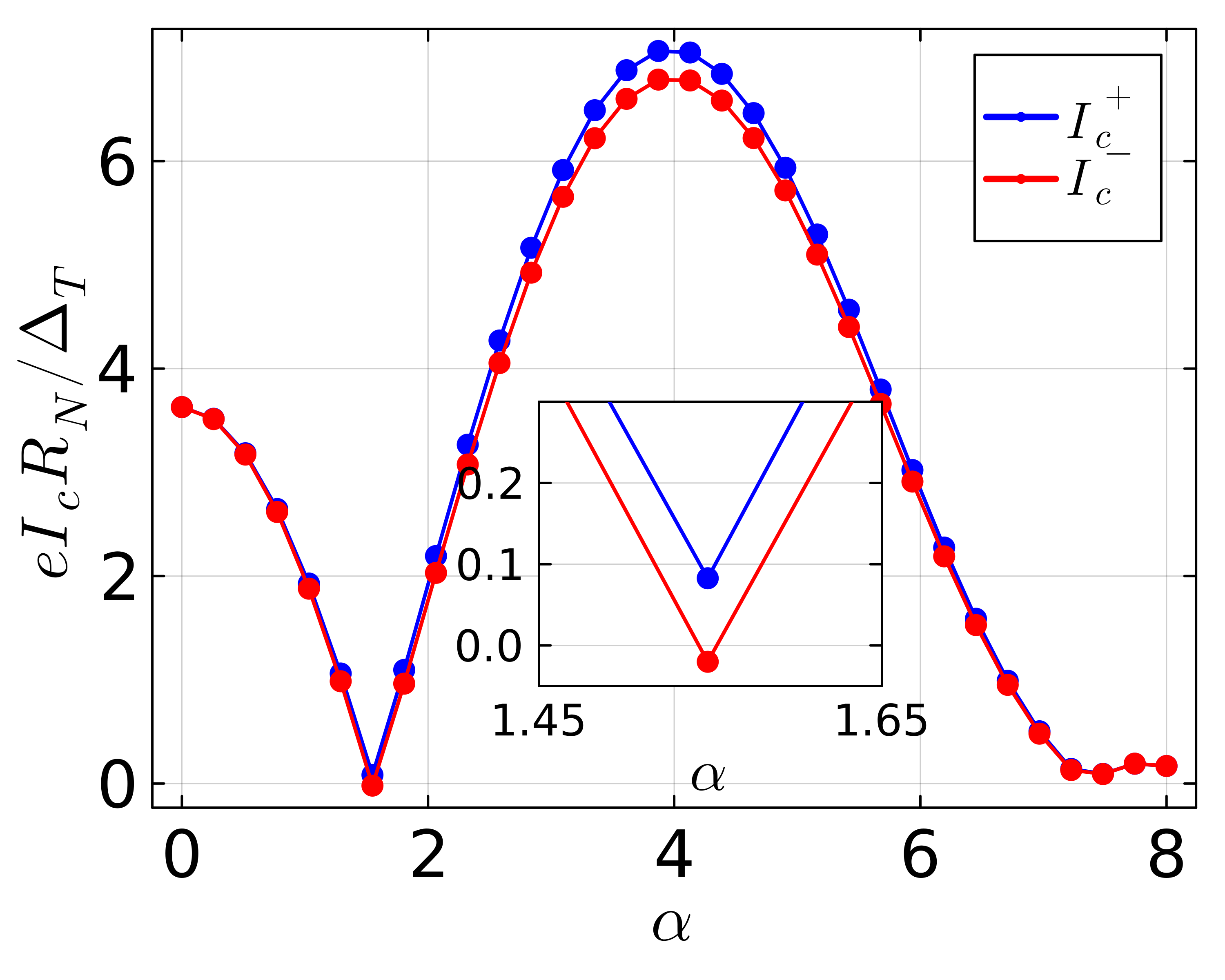}
\end{subfigure}
\begin{subfigure}[b]{0.325\linewidth}
\caption{}\label{subfig:1cprb}
\includegraphics[width=0.999\columnwidth]{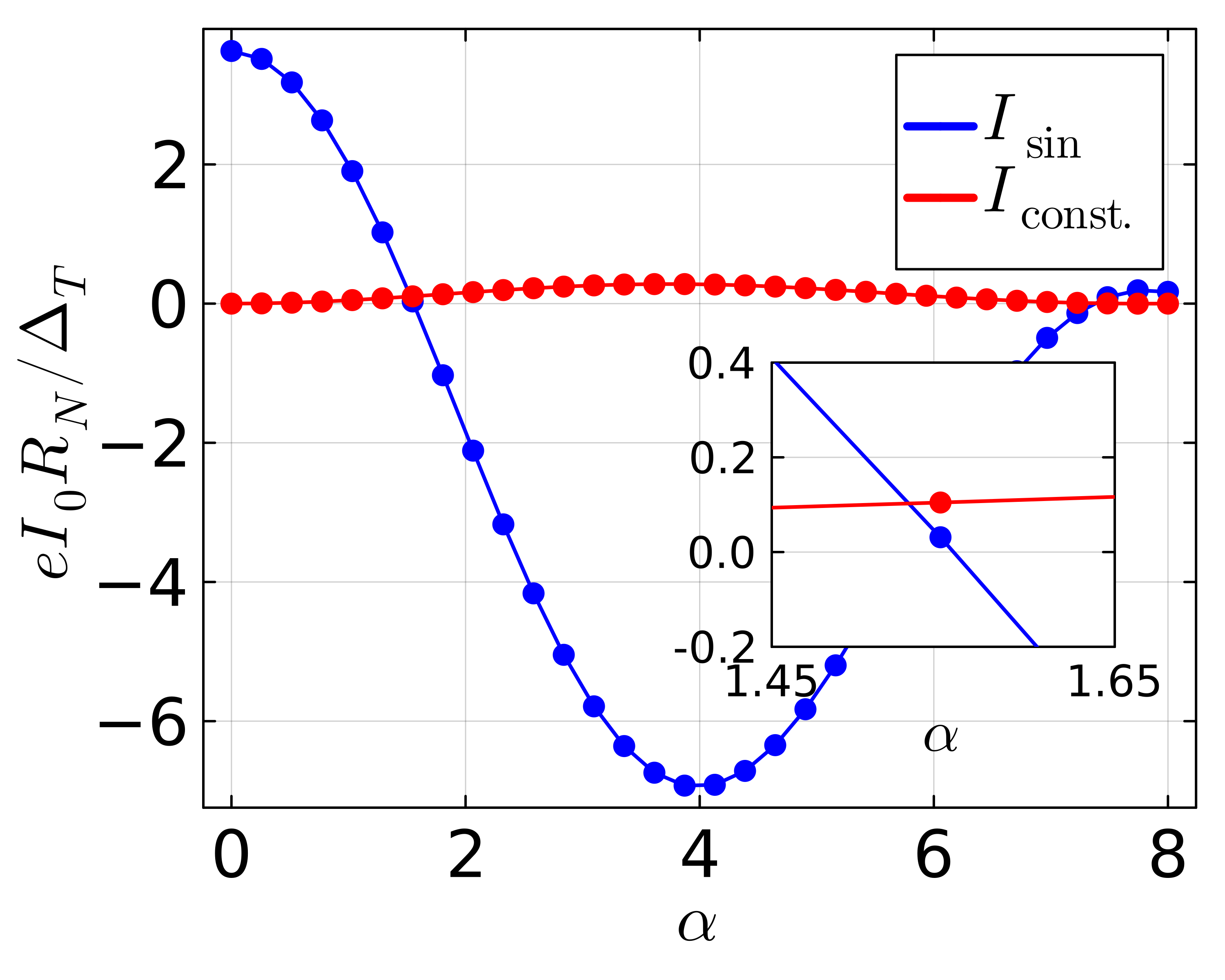}
\end{subfigure}
\caption{Numerically obtained current for the same parameters as in Fig.~\ref{Fig2}(d-f), but with varying $\alpha$ and a fixed radiation frequency $\omega_r=0.3$, close to the YSR transition resonance at $\omega_r=0.368$ (black dashed line in Fig.~\ref{Fig2}(d-f)). The currents are normalized by the numerically obtained normal-state resistance $R_N$. (a) shows the radiation-dressed CPR. For $\alpha=1.548$, $I_{\text{con}}$ is slightly larger than $I_{\sin}$, almost resulting in a perfect diode with $I_c^-\approx 0$. (b) shows $I_c^{\pm}$. We define $I_c^+=\text{max.}[I(\phi)]$ and $I_c^-=-\text{min.}[I(\phi)]$. The sharp corners occur when $I_{\sin}(\alpha)$ vanishes, which causes one of the two critical currents, $I_c^{\pm}$, to vanish, depending on the sign of $I_{\text{con}}$. E.g., when $I_{\text{con}}<0$ and $I_{\sin}=0$, $I_c^+=0$ whereas $I_c^-=|I_{\text{con}}|$. The inset zooms into the region where a perfect diode is obtained, with $I_c^-=0$. (c) shows $I_{\sin}$ and $I_{\text{con}}$ as a function of $\alpha$. The point where the cross over yields a perfect diode.}
\label{Fig3} 
\end{figure*}

\FloatBarrier
$x-z-$plane. A unitary transformation~\cite{Ohnmacht2023} may be performed on the spin-degrees of freedom of both leads to rotate both $\mathbf{J}_T$ and $\mathbf{J}_S$ by an angle $\theta/2$ about the axis $\hat{n}=\mathbf{J}_T\times \mathbf{J}_S/|\mathbf{J}_T\times \mathbf{J}_S|$ which is perpendicular to both $\mathbf{J}_T$ and $\mathbf{J}_S$. Specifically, considering the spin-exchange term in the Hamiltonian, this transformation is given by 
\begin{subequations}
\begin{align}
&U_1=\text{diag}[D_1,\ D_1^*],\quad D_1=\exp(-i\hat{n}\cdot\boldsymbol{\sigma}\theta/4),\\
&U_1^\dagger \text{diag} \big[ \mathbf{J}_{T/S}\cdot\mathbf{\sigma},\ -\big(\mathbf{J}_{T/S}\cdot\mathbf{\sigma}\big)^T\big]U_1 \nonumber\\
 &\hspace{0.5cm}= \ \text{diag}[  D_1^\dagger\ \mathbf{J}_{T/S}\cdot\mathbf{\sigma}\ D_1,-\big(D_1^\dagger\ \mathbf{J}_{T/S}\cdot\mathbf{\sigma}\ D_1\big)^T]\nonumber\\
 &\hspace{0.5cm}= \ \text{diag}[  \mathbf{J}'_{T/S}\cdot\mathbf{\sigma},-\mathbf{J}'_{T/S}\cdot\mathbf{\sigma}\big)^T]\\
&\mathbf{J}'_{T/S}\cdot\mathbf{\sigma}=\big( \mathbf{J}_{T/S}\cos(\theta/2)+(\hat{n}\times \mathbf{J}_{T/S})\sin(\theta/2)\big)\cdot\boldsymbol{\sigma}.
\end{align}
\end{subequations}
We note that it keeps the pairing terms invariant
\begin{align}
D_1^\dagger \Delta i\sigma_y D_1^*=&\ \Delta i\sigma_y.
\end{align}
As a result, both $\mathbf{J}_T$ and $\mathbf{J}_S$ now make an angle $\theta/2$ with the $z-$axis, albeit in opposite directions. Subsequently, another unitary transformation, $U_2=\text{diag}[D_2,\ D_2^*]$ where $D_2=\exp(-i\sigma_z\theta'/2)$, may be performed on both leads about the $z-$axis to align both $\mathbf{J}_T$ and $\mathbf{J}_S$ on the $x-z-$plane, resulting in $\mathbf{J}_T=J[-\sin(\theta/2),\ 0,\ \cos(\theta/2)]$ and $\mathbf{J}_S=J[\sin(\theta/2),\ 0,\ \cos(\theta/2)]$. Since in both steps we perform the same transformation on both leads, the tunnel Hamiltonian remains unchanged~\cite{Ohnmacht2023}. This procedure reflects the fact that in the absence of spin-orbit coupling, there is no absolute reference frame for the spins. We are free to perform a global $\text{SU}(2)$ rotation of the spins to show that the JJ is inversion symmetric, and hence there is no DE. Indeed, Ref.~\cite{Huang2021}, which obtains a closed-form expression for the IVC as a function of the relative angle, does not find a non-reciprocal response. 

We now start by exploring the current as a function of the radiation frequency $\omega_r$ in Fig.~\ref{Fig2}, looking at the impact of non-zero potential scattering, and inversion symmetry breaking. We choose a small value of $\alpha$ because, as seen from the Tien-Gordon expression in Eq.~\eqref{ITG}, it allows us to clearly resolve the role of $\omega_r$. For such small $\alpha$, only the first harmonic of $\omega_r$ contributes appreciably, since higher-order harmonics are strongly suppressed by the decay of the Bessel-function prefactors in Eq.~\eqref{ITG}. We consider larger values of $\alpha$ below in Fig.~\ref{Fig3}, which yields a significantly more pronounced DE. We do not show the trivial case with $K_T=K_S=0$ for brevity, as it clearly precludes a DE. In Fig.~\ref{Fig2}(a-c) we have $K_T=K_S\neq 0$, and while $\mathbf{J}_S$ and $\mathbf{J}_T$ have equal magnitudes, they are misaligned by an angle $\pi/2$. As mentioned earlier, in the absence of spin-orbit coupilng, there is no DE irrespective of the orientation of $\mathbf{J}_T$ and $\mathbf{J}_S$. Now, in Fig.~\ref{Fig2}(d-f), both PHN and inversion symmetries are broken as we have $K_T\neq K_S$. Consequently, as shown in Fig.~\ref{Fig2}(d), we obtain a DE with $I_c^+\neq I_c^-$. This is explained by a non-zero phase-independent current, as show in Fig.~\ref{Fig2}(e), which eventually stems from a non-reciprocal IVC as shown in Fig.~\ref{Fig2}(f). Finally, in Fig.~\ref{Fig2}(g-i), we see a DE as inversion symmetry is broken by $|\mathbf{J}_S|=|\mathbf{J}_T|$, while $K_T=K_S\neq 0$ accounts for PHN breaking.

\begin{figure*}
\begin{subfigure}[b]{0.325\linewidth}
\caption{$\ I_{\text{con}}eR_N/\Delta_T$}\label{subfig:1cprb}
\includegraphics[width=0.999\columnwidth]{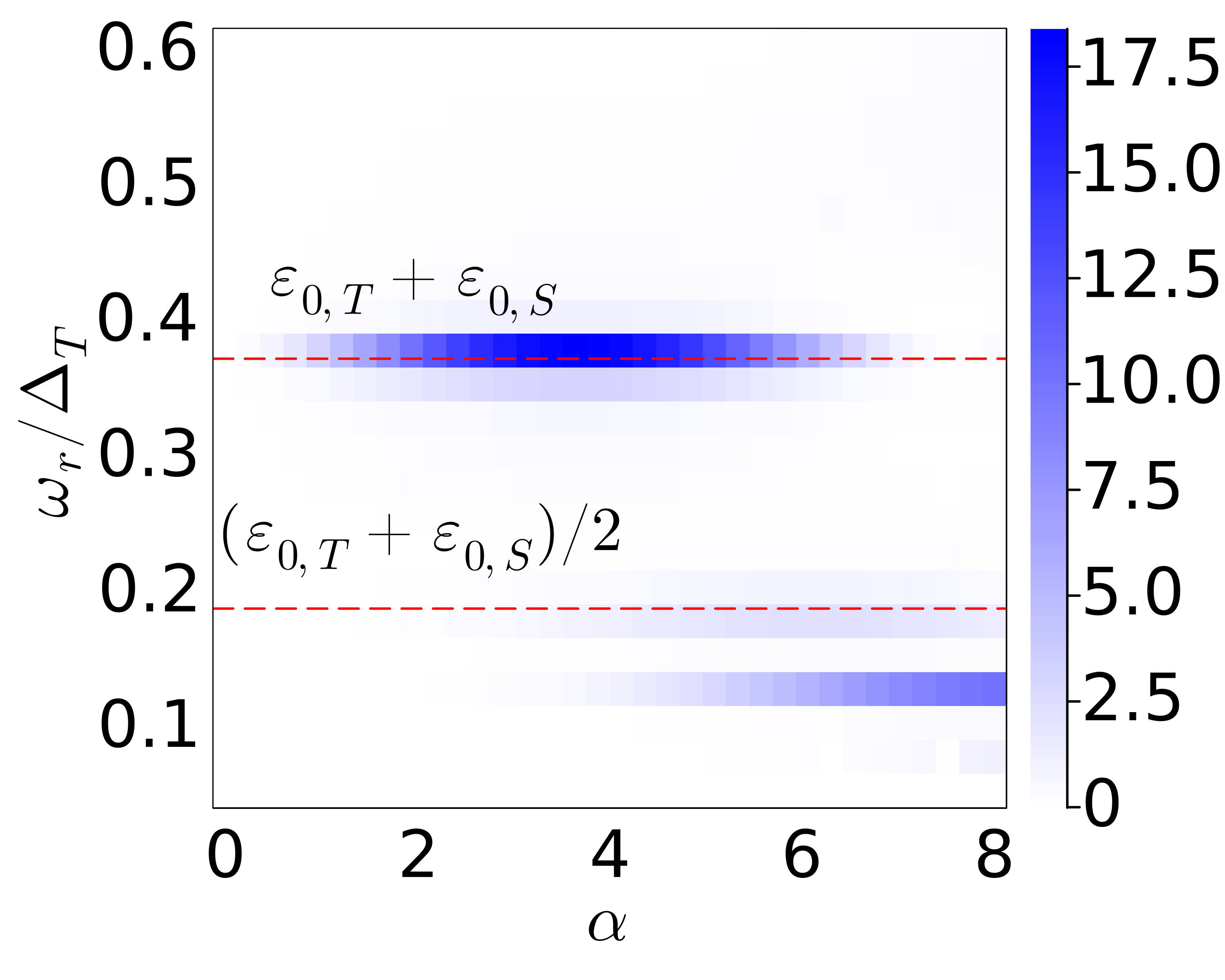}
\end{subfigure}
\begin{subfigure}[b]{0.325\linewidth}
\caption{$\ I_{\sin}eR_N/\Delta_T$}\label{subfig:1cprb}
\includegraphics[width=0.999\columnwidth]{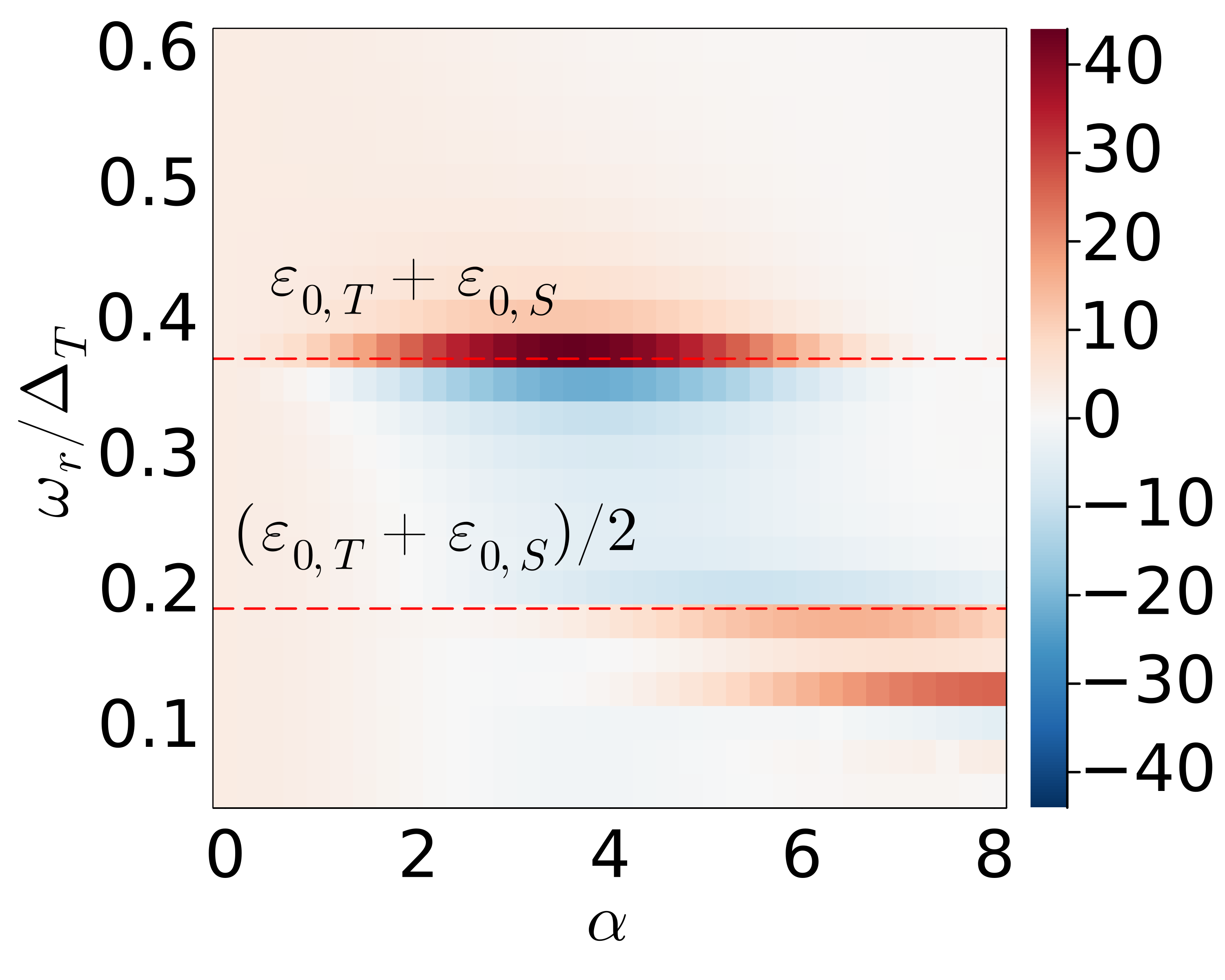}
\end{subfigure}
\begin{subfigure}[b]{0.325\linewidth}
\caption{$\ \eta_{\text{DE}}$}\label{subfig:1cprb}
\includegraphics[width=0.999\columnwidth]{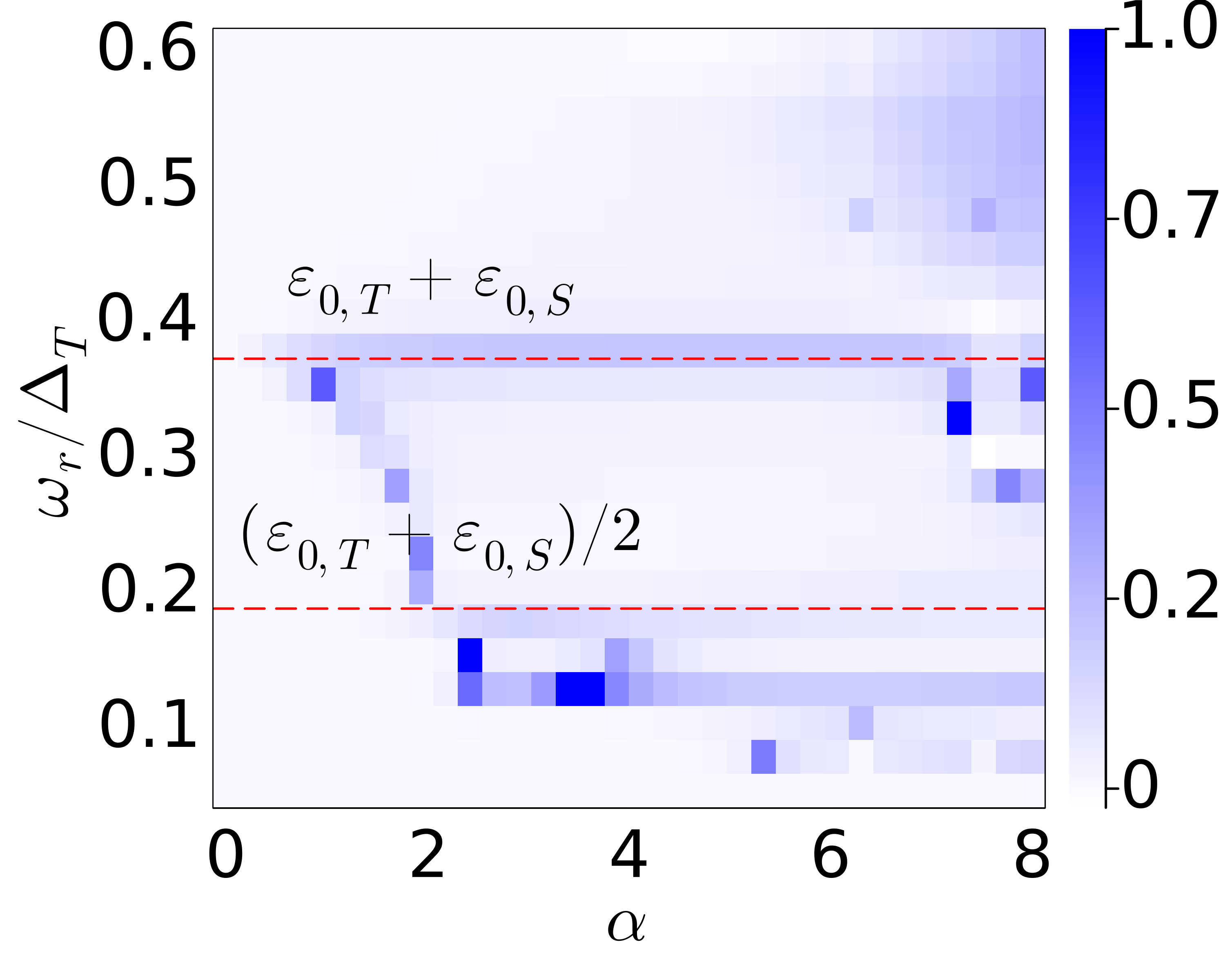}
\end{subfigure}
\caption{Numerically obtained current for the same parameters as in Fig.~\ref{Fig2}(d-f), but with varying $\alpha$ as well as radiation frequency. The currents are normalized by the numerically obtained normal-state resistance $R_N$. In all plots, we denote $\omega_r=\epsilon_{0,T}+\epsilon_{0,S}$ and $\omega_r=(\epsilon_{0,T}+\epsilon_{0,S})/2$, corresponding to the two leading resonances (cf. Eq.~\eqref{ITG}) arising from the YSR transition at $\epsilon_{0,T}+\epsilon_{0,S}$, as shown in Fig.~\ref{Fig2}(d-f). (a) shows $I_{\text{con}}$, revealing the dominant peak at $\omega_r=\epsilon_{0,T}+\epsilon_{0,S}$ for small $\alpha$, followed by a subdominant peak at $\omega_r=(\epsilon_{0,T}+\epsilon_{0,S})/2$ for larger values of $\alpha$. This is because with increasing alpha, the higher order Bessel prefactors in Eq.~\eqref{ITG} start contributing. (b) shows $I_{\sin}$, which is revealing a strong microwave-renormalization with the same resonant features as in (a). The oscillatory dependence on $\alpha$ depends strongly on $\omega_r$, and deviates substantially from the AA ($\sim J_0(\alpha)$). (c) Diode efficiency $\eta_{\text{DE}}$. We obtain high values where $I_{\text{con}}$ becomes non-zero and large enough to match $I_{\sin}$ in magnitude, typically when $I_{\sin}$ nears zero while changing signs as it oscillates as a function of $\alpha$. Remarkably, we even achieve $\eta_{\text{DE}}=1$, a \emph{perfect} diode, near the resonances at $\omega_r=\epsilon_{0,T}+\epsilon_{0,S}$ and $(\epsilon_{0,T}+\epsilon_{0,S})/2$. }
\label{Fig4} 
\end{figure*}

Next, in Fig.~\ref{Fig3}, we investigate how the current depends on the radiation strength $\alpha$ at a fixed, nonzero radiation frequency. The parameters are the same as those used in Fig.~\ref{Fig2}(d-f). Figure~\ref{Fig3}(a) shows the phase dependence of the net dc current $I_0$ [see Eq.~\eqref{If}]. In the absence of irradiation ($\alpha=0$), the CPR is centered at zero and takes the usual sinusoidal form, $I_0(\alpha=0)=I_{\sin}(\alpha=0)\sin(\phi)$, with equal critical currents $I_c^+=I_c^-=I_{\sin}$. However, as $\alpha$ is increased, not only does the amplitude of the CPR vary, but the CPR is also shifted vertically such that it is no longer centered at zero. In other words, $I_0(\alpha)=I_{\sin}(\alpha)\sin(\phi)+I_{\text{con}}(\alpha)$. This shift may even precisely match $I_{\sin}(\alpha)$, such that the critical current vanishes for one polarity, yielding a perfect diode. In Fig.~\ref{Fig3}(a), for $\alpha=1.548$, we are close to this case. This is illustrated in Fig.~\ref{Fig1}(b), where we show the effective critical currents $I_c^+=I_{\sin}(\alpha)+I_{\text{con}}(\alpha)$, and $I_c^-=I_{\sin}(\alpha)-I_{\text{con}}(\alpha)$. As expected, $I_c^+=I_c^-$ at $\alpha=0$. With increasing $\alpha$, they are no longer equal, implying a DE. We zoom into the point exhibiting a perfect DE, where $I_c^-$ vanishes. These results suggest a significantly tunable diode efficiency
\begin{align}
\eta_{\text{DE}}=&\ \frac{||I_c^+|-|I_c^-||}{|I_c^+|+|I_c^-|}\nonumber\\
=&\ \frac{||I_{\sin}(\alpha)+I_{\text{con}}(\alpha)|-|I_{\sin}(\alpha)-I_{\text{con}}(\alpha)||}{|I_{\sin}(\alpha)+I_{\text{con}}(\alpha)|+|I_{\sin}(\alpha)-I_{\text{con}}(\alpha)|}.
\end{align}
In particular, when $|I_{\text{con}}(\alpha)|=|I_{\sin}(\alpha)|$, the efficiency reaches $\eta_{\text{DE}}=1$, corresponding to a perfect diode. This condition is typically accessible. This is because, as illustrated in Fig.~\ref{Fig3}(c), both $I_{\sin}(\alpha)$ and $I_{\text{con}}(\alpha)$ oscillate with $\alpha$. Specifically, $I_{\sin}(\alpha)$ begins at a finite value when $\alpha = 0$ and subsequently changes sign periodically, whereas $I_{\text{con}}(\alpha)$ vanishes at $\alpha = 0$ before developing its oscillatory behavior. A closed-form expression of their functional forms lie beyond the scope of the AA, as mentioned earlier. Additionally, we note that we also obtain situations where $|I_{\text{con}}|>|I_{\sin}|$ and the entire CPR is shifted above(below) zero, amounting to $I_c^-(I_c^+)<0$. Indeed, the curve showing the CPR for $\alpha=1.548$ in Fig.~\ref{Fig3}(a) is one such case with $I_c^-<0$

We now outline practical guidelines for achieving the ideal diode efficiency $\eta_{\text{DE}}=1$, which can be understood as a two-step procedure. First, we recall from the Tien-Gordon expression, Eq.~\eqref{ITG}, that the microwave-induced, phase-independent contribution $I_{\text{con}}(\alpha)$ originates from a nonreciprocal $I_{\text{dc},V}$ evaluated at the voltages $\pm n\omega_r$. As shown in
Fig.~\ref{Fig3}, $I_{\text{dc},V}$ exhibits strong peaks and the largest nonreciprocity at the YSR resonances $eV=|\epsilon_{0,T}\pm\epsilon_{0,S}|$. This motivates choosing radiation frequencies satisfying $\omega_r=|\epsilon_{0,T}\pm\epsilon_{0,S}|/n$. Which of these two resonances appears in the IVC depends on the relative orientation of the magnetic moments. We only give a brief account here; for a detailed discussion we refer to Refs.~\cite{Huang2021,Kleesthesis}. When the two moments are perfectly aligned, only the ``thermal'' resonance at $eV=|\epsilon_{0,T}-\epsilon_{0,S}|$ survives. Its magnitude, however, is strongly suppressed as the temperature is lowered and/or the broadening
$\Gamma$ decreases. In contrast, the ``direct'' resonance at $eV=\epsilon_{0,T}+\epsilon_{0,S}$, which does not suffer from thermal suppression, is forbidden for perfectly aligned spins but becomes allowed for generically misaligned moments. Second, the radiation strength $\alpha$ can be tuned such that $|I_{\text{con}}(\alpha)|=|I_{\sin}(\alpha)|$, which yields a perfect diode. When $\omega_r$ satisfies the resonance conditions discussed above, the AA breaks down. In this regime, neither $I_{\text{con}}(\alpha)$ nor $I_{\sin}(\alpha)$ admits a simple closed-form expression. Nevertheless, our numerical results show that both quantities exhibit oscillatory behavior as functions of $\alpha$, with $I_{\sin}(\alpha=0)=I_c$, while $I_{\text{con}}(\alpha=0)=0$. As a result, there typically exist values of $\alpha$ where´ the condition for a perfect diode is fulfilled. Fig.~\ref{Fig4} illustrates this phenomena, showing $I_{\text{con}}$, $I_{\sin}$ and $\eta_{\text{DE}}$ as a function of $\alpha$ and $\omega_r$ for the same parameters as in Fig.~\ref{Fig2}(d-f). Fig.~\ref{Fig4}(a) shows $I_{\text{con}}$, which is dominated by the fundamental resonance at $\omega_r=\epsilon_{0,T}+\epsilon_{0,S}$ for small $\alpha$, while a subdominant subharmonic resonance emerges at $\omega_r=(\epsilon_{0,T}+\epsilon_{0,S})/2$ as $\alpha$ increases. This behavior reflects the growing importance of higher-order Bessel contributions in Eq.~\eqref{ITG}. The same resonant structure is visible in $I_{\sin}$ shown in Fig.~\ref{Fig4}(b), demonstrating that the microwave renormalization is pronounced only near the YSR transition resonance. Finally, Fig.~\ref{Fig4}(c) displays the diode efficiency $\eta_{\text{DE}}$, which becomes large in the same parameter regime where $I_{\text{con}}$ is enhanced and comparable to $I_{\sin}$. Notably, near the main resonance at $\omega_r=\epsilon_{0,T}+\epsilon_{0,S}$, we find $\eta_{\text{DE}}\approx 1$, realizing a perfect diode.

\section{Discussion and Conclusions}
\label{discconc}
In this work, we present a diode effect which arises only in microwave-driven JJs with broken inversion and PHN symmetries. We identify junctions with YSR states on both superconducting leads~\cite{Huang2020,Huang2021} as an ideal platform to realize this proposal. The critical currents for opposite polarities depend sensitively on the radiation strength and frequency, and thus this setup provides significant tunability for the diode-efficiency. We emphasize that since distinct YSR states are highly unlikely to be absolutely identical, differences in potential scattering and spin-exchange couplings are not only expected but generic. In particular, exact PHN symmetry would require fine-tuning of impurity parameters, which is highly unlikely in practice, as emphasized in Ref.~\cite{Trahms2023} and supported by studies across a range of magnetic impurities. This asymmetry becomes even more natural when different atomic species are employed as magnetic impurities on the two leads, where no symmetry between parameters is anticipated a priori. Experimental observations are consistent with this picture: for instance, Ref.~\cite{Huang2021}, while not focused on diode effects, reports asymmetric differential conductance in a setup involving two YSR states. More broadly, asymmetric differential conductance appears to be a ubiquitous feature~\cite{Farinacci2018,Kot2020,Peters2020}. Furthermore, Ref.~\cite{Siebrecht2023} reports finite potential scattering values extracted from theoretical fits to microwave excitation data of YSR states. Taken together, these findings provide strong support for the feasibility and physical relevance of our proposal.

While we focus on the case where both leads host magnetic impurities, this assumption is not essential. In contrast to a junction with a magnetic impurity in each lead, characterized by YSR energies $\epsilon_{0,T/S}$, in which the YSR-related resonance in the IVC appears at $|\epsilon_{0,T} \pm \epsilon_{0,S}|$, a junction with a single magnetic impurity exhibits a resonance shifted to $\Delta_{T/S} + \epsilon_{0,S/T}$, depending on which lead hosts the YSR state. Since the bulk of the DE occurs from the enhanced non-reciprocity in the IVC at the YSR resonances, this shift necessitates higher microwave frequencies to access the resonance. In fact, since this proposal does not require time-reversal symmetry breaking, instead requiring only broken inversion and PHN symmetries, even clean BCS superconducting leads may, in principle, be used. In this case, PHN is broken by a Fermi level shifted from the band-center. However, as we show in App.~\ref{appB}, we find in this case that the PHN breaking effects are small. In contrast, the consequencies of PHN breaking are much more pronounced with YSRs as they are sharp resonances in frequency, with highly tunable weights depending on magnetic potential scattering and magnetic moments. We emphasize that superconducting leads are essential for realizing a rectifier. In a junction with normal (non-superconducting) leads, the microwave-induced phase-independent contribution persists. However, in the absence of the phase-dependent Josephson term$\sim\sin(\phi)$, there is no finite current interval $I \in [I_c^-, I_c^+]$ over which the voltage vanishes. Such a zero-voltage window is crucial; in a Josephson diode, currents within $[I_c^-, I_c^+]$ are mapped to zero voltage, while currents outside this range produce a finite voltage, thereby enabling rectification~\cite{Wu2022}. Without this extended zero voltage window, a normal junction cannot function as a rectifier.

Finally, we note that some recent works~\cite{Souto2024,Su2024,Matsuo2025} (see also Ref.~\cite{Shaffer2025} and references therein) have investigated the interplay between microwave irradiation and the DE. In both cases, the diode efficiency is tunable via the microwave power. A key distinction from our proposal is that those works begin with Josephson junctions that already exhibit a DE, with the microwave drive providing additional control. In contrast, in our setup the junction displays a DE \emph{only} under microwave irradiation. Furthermore, Refs.~\cite{Su2024,Matsuo2025} operate in the regime where the impedances of the microwave transmission line and surrounding environment exceed that of the junction, resulting in an ac current drive. Here, we instead consider the opposite limit where the microwave radiation imposes an ac voltage~\cite{Cuevas2002}, which is theoretically convenient and permits a fully microscopic treatment. Lastly, we remark that a recent theoretical work~\cite{Shaffer2025a} examines an SNS geometry in which a phase-independent dissipative current is induced in the normal region under an externally applied voltage bias.

\let\oldaddcontentsline\addcontentsline
\renewcommand{\addcontentsline}[3]{}

\section{Acknowledgements}
We thank La\"etitia Farinacci for fruitful discussions. This work was supported by the W\"urzburg-Dresden Cluster of Excellence ct.qmat, EXC2147, Project ID No. 390858490, the collaborative research center, CRC 1170, Project ID No. 258499086, and the DFG research grant TR 950/10.

\appendix

\section{Josephson \emph{tunnel} current in a microwave-irradiated junction}
\label{appA}
In this appendix, we derive the phase-independent dissipative current, as well as the Josephson $\sin\phi$ and $\cos\phi$ current components, for a Josephson \emph{tunnel} junction. We analyze the current response both under a DC voltage bias, and in the presence of microwave irradiation in the absence of an external bias voltage. Our aim is to demonstrate the emergence of a microwave-induced Josephson diode effect in the latter case, and to clarify its connection to the current response under DC voltage bias. We employ the setup described in Sec.~\ref{micromodel} in the tunnel limit, $\mathcal{T}\ll \zeta,\Delta$. 

We consider an analytical approach with a time-domain formalism~\cite{Werthamer1966,Larkin1967,Baronebook,Lahiri2023}, which lets us specialize the results to both biasing cases. In the tunnel limit, the lesser Green's function becomes
\begin{subequations}
\begin{align}
G^<=&\ g^r\Sigma_{\mathcal{T}}^rg^< + g^<\Sigma_{\mathcal{T}}^ag^a,\\
g^<=&\ g^r\Sigma^<g^a,
\end{align}
\end{subequations}
where all the self-energies are as described in Sec.~\ref{micromodel}, and the lowercase Green's functions represent the bare Green's functions of the apexes of the two superconducting leads. In time domain, the products represent convolutions. I.e., for two functions $a\equiv a(t,t')$ and $b\equiv b(t,t')$, the product $ab\equiv [ab](t,t')=\int dt'' a(t,t'')b(t'',t')$. Additionally, the Green's functions are matrices in lead and Nambu space. We suppress the corresponding indices for brevity.

The current is obtained as
\begin{widetext}
\begin{align}
I=&\  e\mathbf{tr}\big[ (\tau_3\otimes \sigma_0)\Sigma_{\mathcal{T},TS}G^<_{ST} \big] - (T\leftrightarrow S)\\
=&\ 2e\mathcal{T}^2\int_{-\infty}^t dt' e^{-i\frac{\phi(t)-\phi(t')}{2}}\Bigg[ \sum_{\eta=1,2,\eta'=1,2}\big( g^>_{T,\eta\eta'}(t,t')g^<_{S,\eta'\eta}(t',t)-g^<_{T,\eta\eta'}(t,t')g^>_{S,\eta'\eta}(t',t) \big) \Bigg]\nonumber\\
&\hspace{1.92cm}- e^{i\frac{\phi(t)-\phi(t')}{2}}\Bigg[ \sum_{\eta=3,4,\eta'=3,4}\big( g^>_{T,\eta\eta'}(t,t')g^<_{S,\eta'\eta}(t',t)-g^<_{T,\eta\eta'}(t,t')g^>_{S,\eta'\eta}(t',t) \big) \Bigg]\nonumber\\
&\hspace{1.72cm}- e^{-i\frac{\phi(t)+\phi(t')}{2}} \Bigg[ \sum_{\eta=1,2} \big( g^>_{T,\eta(5-\eta)}(t,t')g^<_{S,(5-\eta)\eta}(t',t)-g^<_{T,\eta(5-\eta)}(t,t')g^>_{S,(5-\eta)\eta}(t',t) \big) \Bigg]\nonumber\\
&\hspace{1.94cm}+ e^{i\frac{\phi(t)+\phi(t')}{2}}\Bigg[ \sum_{\eta=1,2} \big( g^>_{T,(5-\eta)\eta}(t,t')g^<_{S,\eta(5-\eta)}(t',t)-g^<_{T,(5-\eta)\eta}(t,t')g^>_{S,\eta(5-\eta)}(t',t) \big) \Bigg]
\end{align}\label{It}
\end{widetext}
where we employ $g^r(t-t')=-i\Theta(t-t')\big( g^>(t-t')-g^<(t-t') \big)$. The subscripts $\eta,\eta'$ denote the combined particle-hole and spin Nambu indices. Using $e^{-i\phi(t)/2}=\int(d\omega/(2\pi))W(\omega)e^{-i\omega t}$, $g^<(\omega)=-2i\pi A(\omega)f(\omega)$, and $g^>(\omega)=2i\pi A(\omega)\big(1-f(\omega)\big)$, where the spectral function $A(\omega)=(g^r(\omega)-g^a(\omega))/(-2i\pi)$, we obtain
\begin{widetext}
\begin{align}
I =&\ \Re\  2e \int \frac{d\omega}{2\pi} \frac{d\omega'}{2\pi} e^{-i\omega t-i\omega' t}\Bigg[\nonumber\\
&\ W(\omega) W^*(-\omega') \int d\Omega d\delta\omega\ 
i \frac{
\big[ \sum_{\eta=1,2,\eta'=1,2} A_{T,\eta\eta'}\big(\Omega+\frac{\delta\omega}{2}\big) A_{S,\eta'\eta}\big(\Omega-\frac{\delta\omega}{2}\big) \big]
\big[ f\big(\Omega-\frac{\delta\omega}{2}\big) - f\big(\Omega+\frac{\delta\omega}{2}\big) \big]
}{ \omega'-\delta\omega+i\Gamma }
\nonumber\\
-&\ W^*(-\omega) W(\omega') \int d\Omega d\delta\omega\ 
i \frac{
\big[ \sum_{\eta=3,4,\eta'=3,4} A_{T,\eta\eta'}\big(\Omega+\frac{\delta\omega}{2}\big) A_{S,\eta'\eta}\big(\Omega-\frac{\delta\omega}{2}\big) \big]
\big[ f\big(\Omega-\frac{\delta\omega}{2}\big) - f\big(\Omega+\frac{\delta\omega}{2}\big) \big]
}{ \omega'-\delta\omega+i\Gamma }
\nonumber\\
-&\ W(\omega) W(\omega') \int d\Omega d\delta\omega\ 
i \frac{
\big[ \sum_{\eta=1,2} A_{T,\eta(5-\eta)}\big(\Omega+\frac{\delta\omega}{2}\big) A_{S,(5-\eta)\eta}\big(\Omega-\frac{\delta\omega}{2}\big) \big]
\big[ f\big(\Omega-\frac{\delta\omega}{2}\big) - f\big(\Omega+\frac{\delta\omega}{2}\big) \big]
}{ \omega'-\delta\omega+i\Gamma }\nonumber\\
+&\ W^*(-\omega) W^*(-\omega') \int d\Omega d\delta\omega\ 
i \frac{
\big[ \sum_{\eta=1,2} A_{T,(5-\eta)\eta}\big(\Omega+\frac{\delta\omega}{2}\big) A_{S,\eta(5-\eta)}\big(\Omega-\frac{\delta\omega}{2}\big) \big]
\big[ f\big(\Omega-\frac{\delta\omega}{2}\big) - f\big(\Omega+\frac{\delta\omega}{2}\big) \big]
}{ \omega'-\delta\omega+i\Gamma } \Bigg].\label{Iw}
\end{align}
\end{widetext}

\subsection{Spectral function properties}
We state a few properties of $A_{\eta\eta'}$. In the following discussion, we suppress the subscript $T/S$ for brevity.

$\bullet$ Using
\begin{align}
{g^r}^T(\omega)=&\ {(\omega+i\Gamma-H)^{-1}}^T=(\omega+i\Gamma-H^*)^{-1}\nonumber\\
 =&\ {(\omega-i\Gamma-H)^{-1}}^* = {g^a}^*(\omega),
\end{align}
we derive
\begin{align}
	A_{\eta\eta'}(\omega)=& \frac{1}{2\pi i}\left[{g^a_{\eta\eta'}}(\omega)-{g^r_{\eta\eta'}}(\omega)\right],\\
	=& A^*_{\eta'\eta}(\omega),\qquad (\forall \eta,\eta'). \label{Aprop1}
\end{align}

$\bullet$ The particle-hole symmetry in the superconductor sense (PHS), $\hat{P}=\tau_x\otimes\sigma_0\ \mathcal{K}$, is an \emph{antiunitary} operator which \emph{anticommutes} with the Hamiltonian~\cite{Beenakker2015}
\begin{subequations}
\begin{align}
&\hat{P} H_{T/S} \hat{P}^{-1}=-H_{T/S},\\
\implies &g^r_{T/S}(\omega)=-(\tau_x\otimes\sigma_0){g^r_{T/S}(-\omega)}^*(\tau_x\otimes\sigma_0),
\end{align} 
\end{subequations}
where $\mathcal{K}$ stands for complex conjugation. The matrix $(\tau_x\otimes\sigma_0)$ simply performs $\eta\to \eta+ 2$ for $\eta=\{1,2\}$ and $\eta\to \eta- 2$ for $\eta=\{3,4\}$, exchanging the particle and hole indices. E.g., $g^r_{T/S,11}(\omega)=-{g^r}^*_{T/S,33}(-\omega)$. This implies
\begin{subequations}
\begin{align}
A_{T/S,\eta\eta'}(\omega)=&\ A^*_{T/S,(\eta+2)(\eta'+2)}(-\omega)\quad (\eta,\eta'=1,2),\\
A_{T/S,\eta\eta'}(\omega)=&\ A^*_{T/S,(\eta-2)(\eta'-2)}(-\omega)\quad (\eta,\eta'=3,4).\\
A_{T/S,14}(\omega)=&\ A^*_{T/S,32}(-\omega).
\end{align} \label{Aprop2}
\end{subequations}
\\

$\bullet$ The particle-hole symmetry in the normal-metal sense (PHN) is an \emph{antiunitary} operator that \emph{commutes} with the Hamiltonian. The symmetry is present when $\mu=0$ and $K_{T/S}=0$. If the particle and hole blocks of the Hamiltonian are diagonal (corresponding to ${J_x=J_y=0}$),  ${\hat{P}_\mathrm{PHN}=(\tau_x\otimes\sigma_x)\mathcal{K}}$, leading to
\begin{subequations}
\begin{align}
	&\hat{P}H_{T/S}^D\hat{P}^{-1} = H_{T/S}^D,\\
	\Longrightarrow\; & A_{T/S}^D(\omega) =  (\tau_x\otimes\sigma_x)A_{T/S}^{D*}(\omega)(\tau_x\otimes\sigma_x), \label{ADprop}
\end{align} 
\end{subequations}
where $H^D$ is diagonal in particle and hole blocks and $A^D(\omega)$ is the corresponding spectral function. Hamiltonian with a general $\boldsymbol{J}$ can be transformed to the above mentioned form via a unitary transformation ${H_{T/S}^D=VH_{T/S}V^\dag}$. Under this transformation $A_{T/S}^D=VA_{T/S}V^\dag$ which, together with Eq.~\eqref{ADprop} and with the PHS properties in Eq.~\eqref{Aprop2}, lead to~\cite{Salkota1997,Flette1997,Balatsky2006}
\begin{subequations}
\begin{align}
A_{T/S,11}(\omega)=&\ A_{T/S,44}(\omega)=A_{T/S,22}(-\omega),\\
A_{T/S,33}(\omega)=&\ A_{T/S,22}(\omega)=A_{T/S,44}(-\omega),\\
A_{T/S,12}(\omega)=& -A_{T/S,43}(\omega)=-A_{T/S,12}(-\omega)
\end{align} \label{Aprop3}
\end{subequations}

\subsection{DC voltage bias}
First, we consider the case of a DC voltage bias. We have in this case $\phi(t)=\phi+2eVt$, implying $W(\omega)=\mathcal{T}2\pi\delta(\omega-eV)$. Substituting this in Eq.~\eqref{Iw}, we obtain

\begin{widetext}
\begin{subequations}
\begin{align}
I =&\  I_{\text{dc}, V} + \cos(2eVt) I_{\cos, V}+\sin(2eVt)I_{\sin, V},\\
I_{\text{dc}, V}=& \ 2e\mathcal{T}^2 \int d\Omega d\delta\omega\ \Re\ \Bigg[
i \frac{
\big[ \sum_{\eta=1,2,\eta'=1,2} A_{T,\eta\eta'}\big(\Omega+\frac{\delta\omega}{2}\big) A_{S,\eta'\eta}\big(\Omega-\frac{\delta\omega}{2}\big) \big]
\big[ f\big(\Omega-\frac{\delta\omega}{2}\big) - f\big(\Omega+\frac{\delta\omega}{2}\big) \big]
}{ -eV-\delta\omega+i\Gamma }
\nonumber\\
&\hspace{2.65cm}-
i \frac{
\big[ \sum_{\eta=3,4,\eta'=3,4} A_{T,\eta\eta'}\big(\Omega+\frac{\delta\omega}{2}\big) A_{S,\eta'\eta}\big(\Omega-\frac{\delta\omega}{2}\big) \big]
\big[ f\big(\Omega-\frac{\delta\omega}{2}\big) - f\big(\Omega+\frac{\delta\omega}{2}\big) \big]
}{ eV-\delta\omega+i\Gamma }\Bigg],\\
I_{\cos,V}=&\ 2e\mathcal{T}^2 \int d\Omega d\delta\omega\ \Re\Bigg[ 
i \frac{
\big[ \sum_{\eta=1,2} A_{T,(5-\eta)\eta}\big(\Omega+\frac{\delta\omega}{2}\big) A_{S,\eta(5-\eta)}\big(\Omega-\frac{\delta\omega}{2}\big) \big]
\big[ f\big(\Omega-\frac{\delta\omega}{2}\big) - f\big(\Omega+\frac{\delta\omega}{2}\big) \big]
}{ -eV-\delta\omega+i\Gamma }\nonumber\\
&\ \hspace{2.42cm}-
i \frac{
\big[ \sum_{\eta=1,2} A_{T,\eta(5-\eta)}\big(\Omega+\frac{\delta\omega}{2}\big) A_{S,(5-\eta)\eta}\big(\Omega-\frac{\delta\omega}{2}\big) \big]
\big[ f\big(\Omega-\frac{\delta\omega}{2}\big) - f\big(\Omega+\frac{\delta\omega}{2}\big) \big]
}{ eV-\delta\omega+i\Gamma }
 \Bigg],\\
I_{\sin,V}=&\ -2e\mathcal{T}^2 \int d\Omega d\delta\omega\ \Im\Bigg[ 
i \frac{
\big[ \sum_{\eta=1,2} A_{T,\eta(5-\eta)}\big(\Omega+\frac{\delta\omega}{2}\big) A_{S,(5-\eta)\eta}\big(\Omega-\frac{\delta\omega}{2}\big) \big]
\big[ f\big(\Omega-\frac{\delta\omega}{2}\big) - f\big(\Omega+\frac{\delta\omega}{2}\big) \big]
}{ eV-\delta\omega+i\Gamma }\nonumber\\
&\ \hspace{2.835cm}+i \frac{
\big[ \sum_{\eta=1,2} A_{T,(5-\eta)\eta}\big(\Omega+\frac{\delta\omega}{2}\big) A_{S,\eta(5-\eta)}\big(\Omega-\frac{\delta\omega}{2}\big) \big]
\big[ f\big(\Omega-\frac{\delta\omega}{2}\big) - f\big(\Omega+\frac{\delta\omega}{2}\big) \big]
}{ -eV-\delta\omega+i\Gamma }
 \Bigg].
\end{align}
\end{subequations}
\end{widetext}

Regardless of PHN, when the junction has inversion symmetry (we drop the labels $T,S$ as both leads must be similar)
\begin{widetext}
\begin{align}
I_{\text{dc}, V}\xrightarrow{\text{PHS Eq.}~\eqref{Aprop2}}&\ 2e\mathcal{T}^2 \int d\Omega d\delta\omega\ \Re\ \Bigg[
i \frac{
\big[ \sum_{\eta=1,2,\eta'=1,2} A_{\eta\eta'}\big(\Omega+\frac{\delta\omega}{2}\big) A_{\eta'\eta}\big(\Omega-\frac{\delta\omega}{2}\big) \big]
\big[ f\big(\Omega-\frac{\delta\omega}{2}\big) - f\big(\Omega+\frac{\delta\omega}{2}\big) \big]
}{ -eV-\delta\omega+i\Gamma }
\nonumber\\
&\hspace{2.65cm}-
i \frac{
\big[ \sum_{\eta=1,2,\eta'=1,2} A_{\eta\eta'}\big(-\Omega-\frac{\delta\omega}{2}\big) A_{\eta'\eta}\big(-\Omega+\frac{\delta\omega}{2}\big) \big]
\big[ f\big(\Omega-\frac{\delta\omega}{2}\big) - f\big(\Omega+\frac{\delta\omega}{2}\big) \big]
}{ eV-\delta\omega+i\Gamma }\Bigg]\nonumber\\
\xrightarrow{\Omega\to-\Omega\text{ in }2^{\text{nd}}\text{ line}}&\ 2e\mathcal{T}^2 \int d\Omega d\delta\omega\ \Re\ \Bigg[
i \frac{
\big[ \sum_{\eta=1,2,\eta'=1,2} A_{\eta\eta'}\big(\Omega+\frac{\delta\omega}{2}\big) A_{\eta'\eta}\big(\Omega-\frac{\delta\omega}{2}\big) \big]
\big[ f\big(\Omega-\frac{\delta\omega}{2}\big) - f\big(\Omega+\frac{\delta\omega}{2}\big) \big]
}{ -eV-\delta\omega+i\Gamma }
\nonumber\\
&\hspace{2.65cm}-
i \frac{
\big[ \sum_{\eta=1,2,\eta'=1,2} A_{\eta\eta'}\big(\Omega-\frac{\delta\omega}{2}\big) A_{\eta'\eta}\big(\Omega+\frac{\delta\omega}{2}\big) \big]
\big[ f\big(\Omega-\frac{\delta\omega}{2}\big) - f\big(\Omega+\frac{\delta\omega}{2}\big) \big]
}{ eV-\delta\omega+i\Gamma }\Bigg]\nonumber\\
\xrightarrow{\text{Eq.}~\eqref{Aprop1}}&\ 2e\mathcal{T}^2 \int d\Omega d\delta\omega\ \Bigg[ \sum_{\eta=1,2,\eta'=1,2} A_{\eta\eta'}\bigg(\Omega+\frac{\delta\omega}{2}\bigg) A_{\eta'\eta}\bigg(\Omega-\frac{\delta\omega}{2}\bigg)\Bigg]
\bigg[ f\bigg(\Omega-\frac{\delta\omega}{2}\bigg) - f\bigg(\Omega+\frac{\delta\omega}{2}\bigg) \bigg]\nonumber\\
&\hspace{2.5cm} \Bigg[\frac{\Gamma}{ (eV+\delta\omega)^2+\Gamma^2 }-\frac{\Gamma}{ (-eV+\delta\omega)^2+\Gamma^2 }\Bigg],\quad \implies  I_{\text{dc}, V}(V)=-I_{\text{dc}, V}(-V).
\end{align}
\end{widetext}
In JJs without inversion symmetry ($T$ and $S$ leads are distinct), but having PHN,
\begin{widetext}
\begin{align}
I_{\text{dc}, V}\xrightarrow{\text{PHS Eq.}~\eqref{Aprop2}}&\ 2e\mathcal{T}^2 \int d\Omega d\delta\omega\ \Re\ \Bigg[
i \frac{
\big[ \sum_{\eta=1,2,\eta'=1,2} A_{T,\eta\eta'}\big(\Omega+\frac{\delta\omega}{2}\big) A_{S,\eta'\eta}\big(\Omega-\frac{\delta\omega}{2}\big) \big]
\big[ f\big(\Omega-\frac{\delta\omega}{2}\big) - f\big(\Omega+\frac{\delta\omega}{2}\big) \big]
}{ -eV-\delta\omega+i\Gamma }
\nonumber\\
&\hspace{2.65cm}-
i \frac{
\big[ \sum_{\eta=1,2,\eta'=1,2} A_{T,\eta\eta'}\big(-\Omega-\frac{\delta\omega}{2}\big) A_{S,\eta'\eta}\big(-\Omega+\frac{\delta\omega}{2}\big) \big]
\big[ f\big(\Omega-\frac{\delta\omega}{2}\big) - f\big(\Omega+\frac{\delta\omega}{2}\big) \big]
}{ eV-\delta\omega+i\Gamma }\Bigg]\nonumber\\
\xrightarrow{\text{PHN Eq.}~\eqref{Aprop3}}&\ 2e\mathcal{T}^2 \int d\Omega d\delta\omega\ \Re\Bigg[ \sum_{\eta=1,2,\eta'=1,2} A_{T,\eta\eta'}\bigg(\Omega+\frac{\delta\omega}{2}\bigg) A_{S,\eta'\eta}\bigg(\Omega-\frac{\delta\omega}{2}\bigg)\Bigg]
\bigg[ f\bigg(\Omega-\frac{\delta\omega}{2}\bigg) - f\bigg(\Omega+\frac{\delta\omega}{2}\bigg) \bigg]\nonumber\\
&\hspace{2.75cm} \Bigg[\frac{\Gamma}{ (eV+\delta\omega)^2+\Gamma^2 }-\frac{\Gamma}{ (-eV+\delta\omega)^2+\Gamma^2 }\Bigg],\quad \implies  I_{\text{dc}, V}(V)=-I_{\text{dc}, V}(-V).
\end{align}
\end{widetext} 
The last step, which uses PHN, is crucial to obtain $ I_{\text{dc}, V}(V)=-I_{\text{dc}, V}(-V)$. As seen from the second line in the equation above, the current pathway is mirrored in frequency about the Fermi level for opposite voltage polarities. The presence of PHN makes the amplitudes equal, resulting in $I_{\text{dc}, V}$ being perefctly odd in $V$. Otherwise, in the absence of PHN, these amplitudes are unequal~\cite{Steiner2023,Villas2020} and $I_{\text{dc}, V}$ is not perfectly odd in the voltage $V$, even though it has opposite signs for opposite voltage polarities. We thus conclude that we require broken inversion symmetry as well as broken PHN for the DE.

\begin{figure*}
\begin{subfigure}[b]{0.32\linewidth}
\caption{$\ \mu=0$}\label{subfig:1cprb}
\includegraphics[width=0.999\columnwidth]{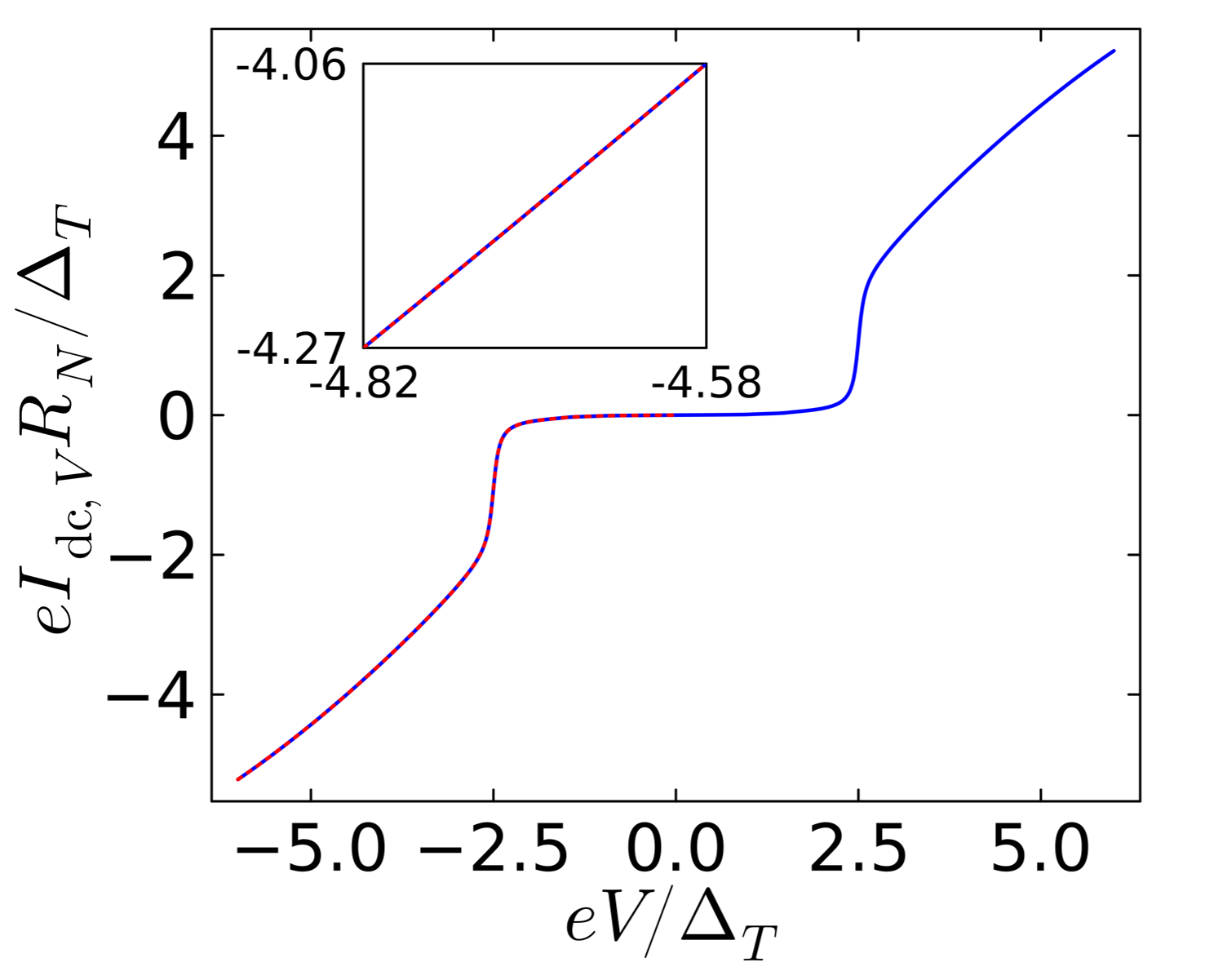}
\end{subfigure}
\begin{subfigure}[b]{0.32\linewidth}
\caption{$\ \mu/\zeta=1.0$}\label{subfig:1cprb}
\includegraphics[width=0.999\columnwidth]{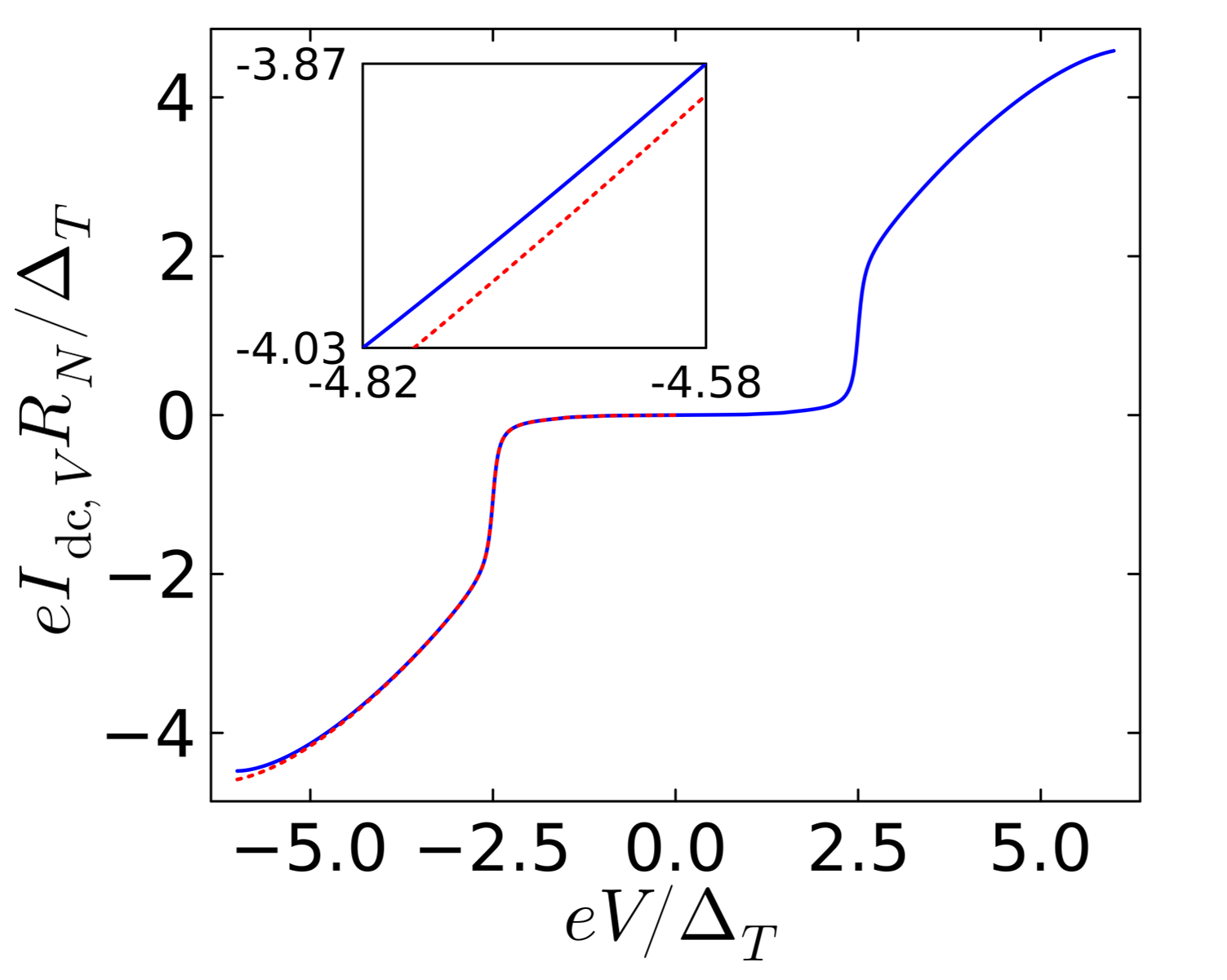}
\end{subfigure}
\begin{subfigure}[b]{0.32\linewidth}
\caption{$\ \mu/\zeta=1.5$}\label{subfig:1cprb}
\includegraphics[width=0.999\columnwidth]{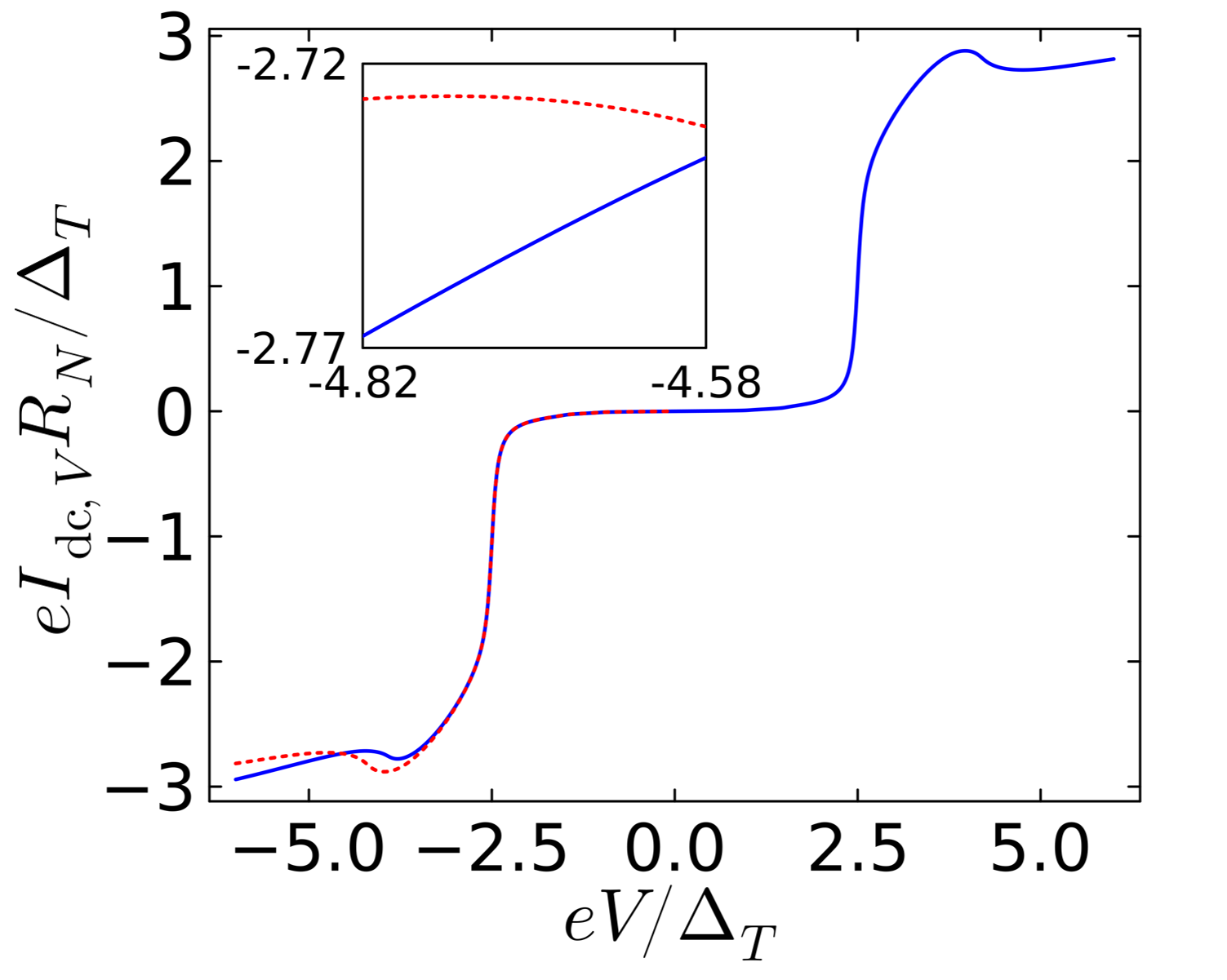}
\end{subfigure}
\caption{Numerically obtained tunnel IVC, $I_{\text{dc},V}$, in a clean tunnel JJ with $\Delta_T=1.0$, $\Delta_S=1.5$, $\Gamma=0.02$, with varying Fermi-level $\mu$ (equal for both leads $\mu_T=\mu_S=\mu$). The bandwidth is $4\zeta$. The currents are normalized by the numerically obtained normal-state resistance $R_N$. For negative voltages, the dotted red line shows $-I_{\text{dc}, V}(-V)$. When it does not match the blue line, the current is non-reciprocal. (a) shows IVC for $\mu=0$, revealing a perfectly odd IVC as we retain PHN symmetry. (b) IVC for $\mu=\zeta$, and (c) IVC for $\mu=1.5\zeta$. They reveal that the non-reciprocity increases commensurately with $\mu$ (see insets for a zoomed-in view). }
\label{Fig5} 
\end{figure*}
Aside, we note that following similar manipulations as above, the amplitudes of the $\cos(2eVt)$ term is found to be perfectly odd in $V$ irrespective of PHN as, on using Eq.~\eqref{Aprop1}, we have $A_{T/S,\eta(5-\eta)}=A^*_{T/S,(5-\eta)\eta}$ for $\eta=1,2$. Similarly, the amplitude of the $\sin(2eVt)$ term is perfectly even in $V$.

\subsection{Microwave irradiated junction with no external voltage bias}
We consider a microwave irradiated JJ, and only provide the expression for the DC current. Using Eq.~\eqref{Wmicro} in the first two lines of Eq.~\eqref{Iw}, we obtain the Tien-Gordon result~\cite{Tien1963,Baronebook}
\begin{widetext}
\begin{align}
I_{\text{dc}}=& \ 2e\mathcal{T}^2 \sum_{n=-\infty}^\infty J_n^2\Big(\frac{\alpha}{2}\Big) \int d\Omega d\delta\omega\ \Re\ \Bigg[
i \frac{
\big[ \sum_{\eta=1,2,\eta'=1,2} A_{T,\eta\eta'}\big(\Omega+\frac{\delta\omega}{2}\big) A_{S,\eta'\eta}\big(\Omega-\frac{\delta\omega}{2}\big) \big]
\big[ f\big(\Omega-\frac{\delta\omega}{2}\big) - f\big(\Omega+\frac{\delta\omega}{2}\big) \big]
}{ -n\omega_r-\delta\omega+i\Gamma }
\nonumber\\
&\hspace{3.55cm}-
i \frac{
\big[ \sum_{\eta=3,4,\eta'=3,4} A_{T,\eta\eta'}\big(\Omega+\frac{\delta\omega}{2}\big) A_{S,\eta'\eta}\big(\Omega-\frac{\delta\omega}{2}\big) \big]
\big[ f\big(\Omega-\frac{\delta\omega}{2}\big) - f\big(\Omega+\frac{\delta\omega}{2}\big) \big]
}{ n\omega_r-\delta\omega+i\Gamma }\Bigg],\\
=& \ \sum_{n=-\infty}^\infty J_n^2\Big(\frac{\alpha}{2}\Big) I_{\text{dc}, V}(n\omega_r).
\end{align}
\end{widetext}
Eq.~\eqref{ITG} follows immediately from this.

\section{Clean BCS superconducting Josephson junction}
\label{appB}
As discussed in Ref.~\cite{Steiner2023}, a non-reciprocity in the $I_{\text{dc},V}$ requires PHN symmetry breaking, and has no relation to time-reversal symmetry breaking. Here, we present the results for clean BCS Josephson junctions with no YSR states. PHN breaking is introduced by a non-zero Fermi level which, as mentioned earlier, results in different density of states for electron- and hole-like excitations. Additionally, inversion symmetry is broken by unequal gaps, $\Delta_T\neq\Delta_S$. We show $I_{\text{dc},V}$ in Fig.~\ref{Fig5}, which shows that the non-reciprocity increases commensurately with chemical potential (difference from the center of the band). Nevertheless, we notice that the non-reciprocity is extremely weak; for $\mu=\zeta$ (bandwidth equals $4\zeta$) it is lower than $1\%$ of the actual magnitude of the current for voltages much larger than $2\Delta$. Furthermore, we find that the non-reciprocity gets weaker with decreasing broadening $\Gamma$. This is because the asymmetrie\deleted{s} in the spectral function introduced by PHN symmetry breaking (cf. Eq.\eqref{Aprop3}) is rather small. Consequently, from Eq.~\eqref{ITG}, we expect negligible DE. 
\FloatBarrier


\begin{thebibliography}{70}

\bibitem{Hu2022} J. Hu, C. Wu, and X. Dai, Proposed design of a Josephson diode, Phys. Rev. Lett. {\bf99}, 067004 (2007).

\bibitem{Chen2018} C.-Z. Chen, J. J. He, M. N. Ali, G.-H. Lee, K. C. Fong, and K. T. Law, Asymmetric josephson effect in inversion symmetry breaking topological materials, Phys. Rev. B {\bf98}, 075430 (2018).

\bibitem{Davydova2022}  M. Davydova, S. Prembabu, and L. Fu, Universal Josephson diode effect, Sci. Adv. {\bf8}, eabo0309 (2022).

\bibitem{Zhang2022}  Y. Zhang, Y. Gu, P. Li, J. Hu, and K. Jiang, General theory of Josephson diodes, Phys. Rev. X {\bf12}, 041013 (2022).

\bibitem{Nadeem2022}  M. Nadeem, M. S. Fuhrer, and X. Wang, The superconducting diode effect, Nat. Rev. Phys. {\bf5}, 558 (2023).

\bibitem{He2022} J. J. He, Y. Tanaka, and N. Nagaosa, A phenomenological theory of superconductor diodes, New J. Phys. {\bf24}, 053014 (2022).

\bibitem{Misaki2021}  K. Misaki and N. Nagaosa, Theory of the nonreciprocal Josephson effect, Phys. Rev. B {\bf103}, 245302 (2021).

\bibitem{Tanaka2022}  Y. Tanaka, B. Lu, and N. Nagaosa, Theory of giant diode effect in d-wave superconductor junctions on the surface of a topological insulator, Phys. Rev. B {\bf106}, 214524 (2022).

\bibitem{Jiang2022}  K. Jiang and J. Hu, Superconducting diode effects, Nat. Phys. {\bf18}, 1145 (2022).

\bibitem{Wang2025} D. Wang, Q.-H. Wang, and C. Wu, Current reversion symmetry breaking and the dc Josephson diode effect, Science Bulletin {\bf70}, 24, 4181 (2025).

\bibitem{Wang2025a} D. Wang, Q.-H. Wang, and C. Wu, Josephson diode effect: a phenomenological perspective, arXiv e-prints, arXiv:2506.23200 (2025).

\bibitem{Wang2025b} R. Wang and N. Hao, Universal diagnostic criterion for intrinsic superconducting diode effect, arXiv e-prints, arXiv:2507.04876 (2025).

\bibitem{Baumgartner2021} C. Baumgartner, L. Fuchs, A. Costa, S. Reinhardt, S. Gronin, G. C. Gardner, T. Lindemann, M. J. Manfra, P. E. F. Junior, D. Kochan, J. Fabian, N. Paradiso, and C. Strunk, Supercurrent rectification and magnetochiral effects in symmetric Josephson junctions, Nat. Nanotechnol. {\bf17}, 39 (2021).

\bibitem{Pal2022} B. Pal, A. Chakraborty, P. K. Sivakumar, M. Davydova, A. K. Gopi, A. K. Pandeya, J. A. Krieger, Y. Zhang, M. Date, S. Ju, N. Yuan, N. B. M. Schr¨oter, L. Fu, and S. S. P. Parkin, Josephson diode effect from Cooper pair momentum in a topological semimetal, Nat. Phys. {\bf18}, 1228 (2022).

\bibitem{Diezmerida2023} J. D\'iez-M\'erida, A. D\'iez-Carl\'on, S. Y. Yang, Y. M. Xie, X. J. Gao, K. Watanabe, T. Taniguchi, X. Lu, A. P. Higginbotham, K. T. Law, and D. K. Efetov, Symmetry-broken Josephson junctions and superconducting diodes in magicangle twisted bilayer graphene, Nat. Commun. {\bf14}, 2396 (2023).

\bibitem{Wu2022} H. Wu, Y. Wang, Y. Xu, P. K. Sivakumar, C. Pasco, U. Filippozzi, S. S. P. Parkin, Y.-J. Zeng, T. McQueen, and M. N. Ali, The field-free Josephson diode in a van der Waals heterostructure, Nature {\bf604}, 653 (2022).

\bibitem{Bocquillon2017} E. Bocquillon, R. S. Deacon, J. Wiedenmann, P. Leubner, T. M. Klapwijk, C. Br¨une, K. Ishibashi, H. Buhmann, and L. W. Molenkamp, Gapless Andreev bound states in the quantum spin Hall insulator HgTe, Nat. Nanotech. {\bf12}, 137 (2017).

\bibitem{Bauriedl2022} L. Bauriedl, C. B\"auml, L. Fuchs, C. Baumgartner, N. Paulik, J. M. Bauer, K.-Q. Lin, J. M. Lupton, T. Taniguchi, K. Watanabe, C. Strunk, and N. Paradiso, Nat. Commun. {\bf13}, 4266 (2022).

\bibitem{Jeon2022} K.-R. Jeon, J.-K. Kim, J. Yoon, J.-C. Jeon, H. Han, A. Cottet, T. Kontos, and S. S. P. Parkin, Nat. Mat. {\bf21}, 1008 (2022).

\bibitem{Turini2022} B. Turini, S. Salimian, M. Carrega, A. Iorio, E. Strambini, F. Giazotto, V. Zannier, L. Sorba, and S. Heun,
Nano Lett. 22, 8502 (2022).

\bibitem{Gupta2023} M. Gupta, G. V. Graziano, M. Pendharkar, J. T. Dong, C. P. Dempsey, C. Palmstrøm, and V. S. Pribiag, Superconducting diode effect in a three-terminal Josephson device, Nat. Commun. {\bf14}, 3078 (2023).

\bibitem{Chiles2023} J. Chiles, E. G. Arnault, C.-C. Chen, T. F. Q. Larson, L. Zhao, K. Watanabe, T. Taniguchi, F. Amet, and G. Finkelstein, Nonreciprocal supercurrents in a field-free graphene Josephson triode, Nano Lett. 23, 5257 (2023).

\bibitem{Kokkeler2022} T. H. Kokkeler, A. A. Golubov, and F. S. Bergeret, Field-free anomalous junction and superconducting diode effect in spin-split superconductor/topological insulator junctions, Phys. Rev. B {\bf106}, 214504 (2022).

\bibitem{Zhao2023} S. Y. F. Zhao, X. Cui, P. A. Volkov, H. Yoo, S. Lee, J. A. Gardener, A. J. Akey, R. Engelke, Y. Ronen, R. Zhong, G. Gu, S. Plugge, T. Tummuru, M. Kim, M. Franz, J. H. Pixley, N. Poccia, and P. Kim, Time-reversal symmetry breaking superconductivity between twisted cuprate superconductors, Science {\bf382}, 1422 (2023).

\bibitem{Zhang2024} F. Zhang, M. T. Ahari, A. S. Rashid, G. J. de Coster, T. Taniguchi, K. Watanabe, M. J. Gilbert, N. Samarth, and M. Kayyalha, Reconfigurable magnetic-field-free superconducting diode effect in multi-terminal Josephson junctions, Phys. Rev. Appl. {\bf21}, 034011 (2024).

\bibitem{Cheng2024} Q. Cheng, Y. Mao, and Q.-F. Sun, Field-free Josephson diode effect in altermagnet/normal metal/altermagnet junctions, Phys. Rev. B {\bf110}, 014518 (2024).

\bibitem{Banerjee2024} S. Banerjee and M. S. Scheurer, Altermagnetic superconducting diode effect, Phys. Rev. B 110, 024503 (2024).

\bibitem{Ma2025} J. Ma, H. Wang, W. Zhuo, B. Lei, S. Wang, W. Wang, X.- Y. Chen, Z.-Y. Wang, B. Ge, Z. Wang, J. Tao, K. Jiang, Z. Xiang, and X.-H. Chen, Field-free Josephson diode effect in NbSe2 van der Waals junction, Commun. Phys. {\bf8}, 125 (2025).

\bibitem{Nagata2025} U. Nagata, M. Aoki, A. Daido, S. Kasahara, Y. Kasahara, R. Ohshima, Y. Ando, Y. Yanase, Y. Matsuda, and M. Shiraishi, Field-free superconducting diode effect in layered superconductor FeSe, Phys. Rev. Lett. {\bf134}, 236703 (2025).

\bibitem{Trahms2023} M. Trahms, L. Melischek, J. F. Steiner, B. Mahendru, I. Tamir, N. Bogdanoff, O. Peters, G. Reecht, C. B. Winkelmann, F. von Oppen, and K. J. Franke, Diode effect in Josephson junctions with a single magnetic atom, Nature {\bf615}, 628 (2023).

\bibitem{Steiner2023} J. F. Steiner, L. Melischek, M. Trahms, K. J. Franke, and F. von Oppen, Diode effects in current-biased Josephson junctions, Phys. Rev. Lett. {\bf130}, 177002 (2023).

\bibitem{Trahms2025} M. Trahms, B. Mahendru, C. B. Winkelmann, and K. J. Franke, From Shapiro steps to photon-assisted tunneling in microwave-driven atomic-scale Josephson junctions with a single (magnetic) adatom, arXiv e-prints, arXiv:2509.26228 (2025).

\bibitem{Ghosh2024} S. Ghosh, V. Patil, A. Basu, Kuldeep, A. Dutta, D. A. Jangade, R. Kulkarni, A. Thamizhavel, J. F. Steiner, F. von Oppen, and M. M. Deshmukh, High-temperature Josephson diode, Nat. Mater. {\bf23}, 612 (2024).

\bibitem{Yu1965} L. Yu, Bound state in superconductors with paramagnetic impurities, Acta Phys. Sin. {\bf21}, 75 (1965).

\bibitem{Shiba1968} H. Shiba, Classical Spins in Superconductors, Prog. Theor. Phys. {\bf40}, 435 (1968).

\bibitem{Rusinov1969} A. I. Rusinov, Superconductivity near a paramagnetic impurity, JETP Lett. {\bf9}, 85 (1969).

\bibitem{Baronebook} A. Barone and G. Paterno, Physics and Applications of the Josephson Effect (Wiley, New York, 1982).

\bibitem{Cuevas2002} J. C. Cuevas, J. Heurich, A. Mart\'in-Rodero, A. Levy Yeyati, and G. Sch\"on, Subharmonic Shapiro Steps and Assisted Tunneling in Superconducting Point Contacts, Phys. Rev. Lett. {\bf88}, 157001 (2002).

\bibitem{Chauvinthesis} M. Chauvin, The Josephson Effect in Atomic Contacts, Ph.D. thesis, SPEC/CEA-Saclay (2005).

\bibitem{Shapiro1963} S. Shapiro, Josephson Currents in Superconducting Tunneling: The Effect of Microwaves and Other Observations, Phys. Rev. Lett. {\bf11}, 80 (1963).

\bibitem{Kot2020} P. Kot, R. Drost, M. Uhl, J. Ankerhold, J. C. Cuevas, C. R. Ast, Microwave-assisted tunneling and interference effects in superconducting junctions under fast driving signals, Phys. Rev. B. {\bf101}, 13, 134507 (2020).

\bibitem{Siebrecht2023} J. Siebrecht, H. Huang, P. Kot, R. Drost, C. Padurariu, B. Kubala, J. Ankerhold, J. C. Cuevas, and C. R. Ast, Microwave excitation of atomic scale superconducting bound states, Nat. Commun. {\bf14}, 6794 (2023).

\bibitem{Bergeret2010} F. S. Bergeret, P. Virtanen, T. T. Heikkil\"a, and J. C. Cuevas, Theory of Microwave-Assisted Supercurrent in Quantum Point Contacts, Phys. Rev. Lett. {\bf 105}, 117001 (2010). 

\bibitem{Werthamer1966} N. R. Werthamer, Nonlinear Self-Coupling of Josephson Radiation in Superconducting Tunnel Junctions, Phys. Rev. {\bf147}, 255 (1966).

\bibitem{Larkin1967} A. I. Larkin and Y. N. Ovchinnikov, tunnel effect between superconductors in an alternating field, Zh. Eksp. Teor. Fiz. {\bf51}, 1535 [Sov. Phys. JETP {\bf24}, 1035 (1967)].

\bibitem{Huang2021} H. Huang, J. Senkpiel, C. Padurariu, R. Drost, A. Villas, R. L. Klees, A. L. Yeyati, J. C. Cuevas, B. Kubala, J. Ankerhold, K. Kern, and C. R. Ast, Spin-dependent tunneling between individual superconducting bound states, Phys. Rev. Research {\bf3}, L032008 (2021).

\bibitem{Tien1963} P. Tien and J. Gordon, Phys. Rev. {\bf129}, 647 (1963).

\bibitem{Falci1991} G. Falci, V. Bubanja, and G. Sch\"on, Quasiparticle and Cooper pair tunneling in small capacitance Josephson junctions, Z. Phys. B {\bf85}, 451 (1991).

\bibitem{Safi2010} I. Safi and E. V. Sukhorukov, Determination of tunneling charge via current measurements, Europhysics Letters {\bf91}, 67008 (2010).

\bibitem{Safi2019} I. Safi, Driven quantum circuits and conductors: A unifying perturbative approach, Phys. Rev. B {\bf99}, 045101 (2019)

\bibitem{Salkota1997} M. I. Salkola, A. V. Balatsky, J. R. Schrieffer, Phys. Rev. B {\bf55}, 12648 (1997).

\bibitem{Flette1997} M. E. Flatt\'e and J. M. Byers, Local electronic structure of defects in superconductors, Phys. Rev. B {\bf56}, 11213 (1997).

\bibitem{Balatsky2006} A. V. Balatsky, I. Vekhter, and J.-X. Zhu, Impurity-induced states in conventional and unconventional superconductors, Rev. Mod. Phys. {\bf78}, 373 (2006).

\bibitem{Villas2020} A. Villas, R. L. Klees, H. Huang, C. R. Ast, G. Rastelli, W. Belzig, and J. C. Cuevas, Interplay between Yu-Shiba-Rusinov states and multiple andreev reflections, Phys. Rev. B {\bf101}, 235445 (2020).

\bibitem{Cuevas1996} J. C. Cuevas, A. Mart\'{i}n-Rodero, and A. Levy Yeyati, Hamiltonian approach to the transport properties of superconducting quantum point contacts, Phys. Rev. B {\bf54}, 7366 (1996).

\bibitem{Lahiri2025} A. Lahiri, S.-J. Choi, B. Trauzettel, AC Josephson Signatures of the Superconducting Higgs Mode, Phys. Rev. B {\bf112}, 094516 (2025).

\bibitem{Lahiri2026} A. Lahiri, J. C. Cuevas, B. Trauzettel, Signatures of superconducting Higgs mode in irradiated Josephson junctions, Phys. Rev. B {\bf113}, 014516 (2026).

\bibitem{Chakraborty2020} S. Chakraborty, D. Nikoli\'c, R. S. Souto, W. Belzig, and J. C. Cuevas, DC Josephson effect between two Yu-Shiba-Rusinov bound states, Phys. Rev. B {\bf108}, 094518 (2023).

\bibitem{Beenakker2015} C. W. J. Beenakker, Random-matrix theory of Majorana fermions and topological superconductors, Rev. Mod. Phys. {\bf87}, 1037 (2015).

\bibitem{Samanta1998} M. P. Samanta and S. Datta, Electrical transport in junctions between unconventional superconductors: Application of the Green’s-function formalism, Phys. Rev. B {\bf57}, 10972 (1998).

\bibitem{Jauho1994} A.-P. Jauho, N. S. Wingreen, and Y. Meir, Time-dependent transport in interacting and noninteracting resonant-tunneling systems, Phys. Rev. B {\bf50}, 5528 (1994).

\bibitem{Keldysh1964} L. V. Keldysh, Diagram technique for nonequilibrium processes, Sov. Phys. JETP {\bf20}, 1018 (1964).

\bibitem{Stefanuccibook2013} G. Stefanucci and R. van Leeuwen, \textit{Nonequilibrium Many-Body Theory of Quantum Systems: A Modern Introduction}, Cambridge University Press, Cambridge, (2013).

\bibitem{Gonzales2020} S. A. Gonz\'{a}lez, L. Melischek, O. Peters, K. Flensberg, K. J. Franke, and F. von Oppen, Photon-assisted resonant Andreev reflections: Yu-Shiba-Rusinov and Majorana states, Phys. Rev. B {\bf102}, 045413 (2020).

\bibitem{Wimmerthesis} M. Wimmer, Quantum transport in nanostructures: From computational concepts to spintronics in graphene and magnetic tunnel junctions, Ph.D. thesis, Universit\" at Regensburg (2008).

\bibitem{Gldecaynote} $G^<=(1+\Sigma^rG^r)g^<(1+G^a\Sigma^a)+G^r\Sigma^< G^a$. The first term refers to the initial conditions. In the presence of a finite lifetime $\sim 1/\Gamma$, it decays to zero before the bias voltage is applied due to the exponentially decaying $G^{r/a}$~\cite{Keldysh1964,Stefanuccibook2013,Gonzales2020,Wimmerthesis}. 

\bibitem{Ohnmacht2023} D. C. Ohnmacht, W. Belzig, and J. C. Cuevas, Full counting statistics of Yu-Shiba-Rusinov bound states, Phys. Rev. Res. {\bf5}, 033176 (2023).

\bibitem{Kleesthesis} R. Klees, Nonequilibrium Transport and Dynamics in Conventional and Topological Superconducting Junctions, Ph.D. thesis, Universit\"at Konstanz (2021).

\bibitem{tdnote} Even though we have $\theta=0$ and our calculations assume zero temperature, we still observe the (thermal) peak at $|\epsilon_{0,T}-\epsilon_{0,S}|$ in $I_{\text{dc},V}$ due to a finite value of $\Gamma$. For non-zero values of $\theta$ in the same system, we obtain a much larger (direct) peak at $|\epsilon_{0,T}+\epsilon_{0,S}|$.

\bibitem{Huang2020} H. Huang, C. Padurariu, J. Senkpiel, R. Drost, A. Levy Yeyati, J. C. Cuevas, B. Kubala, J. Ankerhold, K. Kern and C. R. Ast, Tunnelling dynamics between superconducting bound states at the atomic limit. Nat. Phys. 16, 1227-1231 (2020).

\bibitem{Farinacci2018} L. Farinacci, G. Ahmadi, G. Reecht, M. Ruby, N. Bogdanoff, O. Peters, B. W. Heinrich, F. von Oppen, and
K. J. Franke, Tuning the Coupling of an Individual Magnetic Impurity to a Superconductor: Quantum Phase Transition and Transport, Phys. Rev. Lett. {\bf121}, 196803 (2018).

\bibitem{Peters2020} O. Peters, N. Bogdanoff, S. Acero Gonz´alez, L. Melischek, J. R. Simon, G. Reecht, C. B. Winkelmann, F. von
Oppen, and K. J. Franke, Resonant Andreev reflections probed by photon-assisted tunnelling at the atomic scale, Nature Physics {\bf16}, 1222 (2020).

\bibitem{Souto2024} R. Seoane Souto, M. Leijnse, C. Schrade, M. Valentini, G. Katsaros, and J. Danon, Tuning the Josephson diode response with an ac current, Phys. Rev. Res. {\bf6}, L022002 (2024).

\bibitem{Su2024} H. Su, J.-Y. Wang, H. Gao, Y. Luo, S. Yan, X. Wu, G. Li, J. Shen, L. Lu, D. Pan, J. Zhao, P. Zhang, and H. Q. Xu, Microwave-Assisted Unidirectional Superconductivity in Al-InAs Nanowire-Al Junctions under Magnetic Fields. Phys. Rev. Lett. {\bf133}, 087001 (2024).

\bibitem{Matsuo2025} S. Matsuo, R. S. Deacon, S. Kobayashi, Y. Sato, T. Yokoyama, T. Lindemann, S. Gronin, G. C. Gardner, K. Ishibashi, M. J. Manfra, and S. Tarucha, Shapiro response of superconducting diode effect derived from Andreev molecules, Phys. Rev. B {\bf111}, 094512 (2025).

\bibitem{Shaffer2025} D. Shaffer, A. Levchenko, Theories of Superconducting Diode Effects, arXiv e-prints, arXiv:2510.25864 (2025).

\bibitem{Shaffer2025a} D. Shaffer, S. Li, J. Hasan, M. Titov, and A. Levchenko, Josephson diode effect from nonequilibrium current in a superconducting interferometer, Phys. Rev. B {\bf112}, 094509 (2025).

\bibitem{Lahiri2023} A. Lahiri, S.-J. Choi, and B. Trauzettel, Nonequilibrium Fractional Josephson Effect, Phys. Rev. Lett. {\bf131}, 126301 (2023).

\end{thebibliography}
\end{document}